\documentclass[sigconf, authorversion,nonacm, 10pt]{acmart}
\usepackage[utf8]{inputenc}
\usepackage[capitalise]{cleveref}
\usepackage{color}
\usepackage{booktabs}
\usepackage{xspace}
\usepackage{amsmath}
\usepackage{siunitx}
\usepackage{enumitem}
\usepackage{tabularx}
\usepackage{xcolor}
\usepackage{mdframed}
\usepackage{graphicx,subfigure}
\usepackage[font=footnotesize,labelfont=bf]{caption}
\usepackage{soul}
\usepackage[most]{tcolorbox}
\usepackage{listings}
\usepackage{verbatim}
\usepackage{verbatimbox}
\usepackage{fancyvrb}

\newcommand{\fakeparagraph}[1]{\vspace{.5mm}\noindent\textbf{#1.}}
\newcommand{\fakepar}[1]{\fakeparagraph{#1}}

\setlist[itemize]{leftmargin=*,noitemsep}
\usepackage[T1]{fontenc}
\newcommand\thefont{\expandafter\string\the\font}

\newenvironment{itemize*}{\begin{itemize}}{\end{itemize}}

\begin{document}

\newcommand{\name}{{\em PixelGen}}
\newcommand{\nameF}{{PixelGen}}
\newcommand{\system}{{PixelGen}}
\newcommand{\systemF}{{PixelGen}}

\newcommand{\hardware}{{PixelSense}}
\newcommand{\hardwareF}{{PixelSense}}

\title{\system: Rethinking Embedded Camera Systems}

\author[K.Li]{Kunjun Li}
\affiliation{
  \institution{National University of Singapore}
  \country{}
}
\email{kunjun_li@u.nus.edu}

\author [M.Gulati]{Manoj Gulati}
\affiliation{
  \institution{National University of Singapore}
  \country{}
}
\email{manojg@nus.edu.sg}

\author [S.Waskito] {Steven Waskito }
\affiliation{
  \institution{National University of Singapore}
  \country{}
}
\email{steven.waskito@u.nus.edu}

\author [D.Shah] {Dhairya Shah}
\affiliation{
  \institution{National University of Singapore}
  \country{}
}
\email{dhairya@nus.edu.sg}

\author[S.Chakrabarty] {Shantanu Chakrabarty}
\affiliation{
  \institution{NCS Group}
  \country{}
}
\email{shantanu.chakrabarty@ncs.com.sg}

\author[A.Varshney] {Ambuj Varshney}
\affiliation{
  \institution{National University of Singapore}
  \country{}
}
\email{ambujv@nus.edu.sg}

\begin{teaserfigure}
	\includegraphics[width=\textwidth]{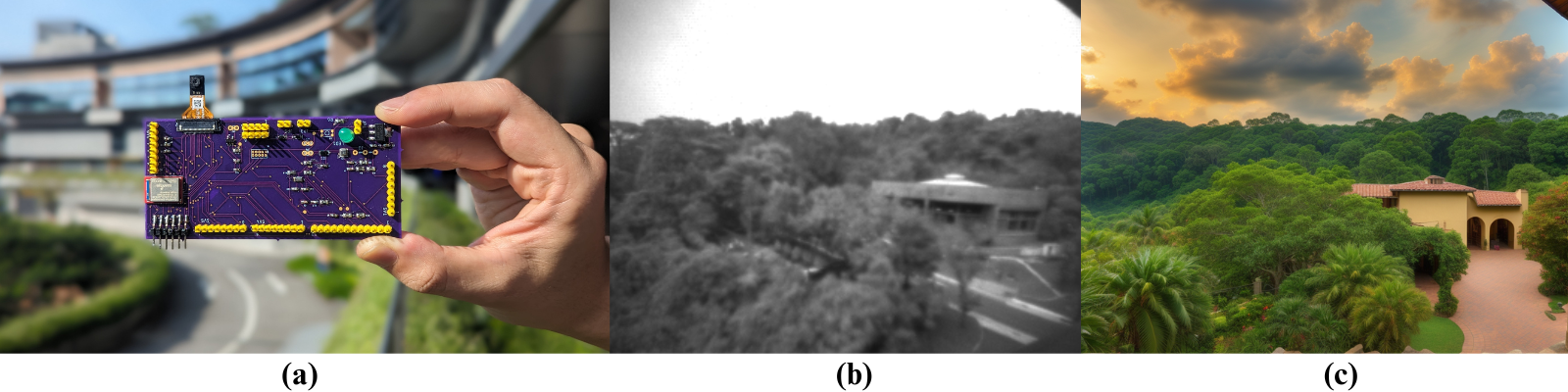}
        \caption{\system\space  rethinks the architecture of an embedded camera systems. \system\space captures a comprehensive representation of the world through an array of simple sensors and  transceivers. They gather environmental information, including temperature, humidity, ambient light conditions, acoustic emissions, motion data, and prevalent radio and magnetic signals. Additionally, a low-resolution image sensor operates in visible and infrared spectrum to capture visual representations of the surroundings. The simplicity of the sensor, and low-resolution imagery~(b), facilitates the design of a low-power hardware platform, as illustrated in (a). This also enables low-bitrate transmissions, and thus minimizing bandwidth usage. The captured data is then processed using a transformer-based image diffusion model, enabling the generation of novel environmental representations. This includes enabling the generating of high-definition images~(c).}
	\label{fig:teaser}
\end{teaserfigure}

\begin{abstract}
Embedded camera systems are ubiquitous, representing the most widely deployed example of a wireless embedded system. They capture a representation of the world — the surroundings illuminated by visible or infrared light.  Despite their widespread usage, the architecture of embedded camera systems has remained unchanged, which leads to limitations. They visualize only a tiny portion of the world. Additionally, they are energy-intensive, leading to limited battery lifespan. We present \systemF,  which re-imagines embedded camera systems. Specifically, \system\space combines sensors, transceivers, and low-resolution image and infrared vision sensors to capture a broader world representation. They are deliberately chosen for their simplicity, low bitrate, and power consumption, culminating in an energy-efficient platform. We show that despite the simplicity, the captured data can be processed using transformer-based image and language models to generate novel representations of the environment. For example, we demonstrate that it can allow the generation of high-definition images, while the camera utilises low-power, low-resolution monochrome cameras. Furthermore, the capabilities of \system\space extend beyond traditional photography, enabling visualization of phenomena invisible to conventional cameras, such as sound waves. \system\space can enable numerous novel applications, and we demonstrate that it enables unique visualization of the surroundings that are then projected on extended reality headsets. We believe, \system\space goes beyond conventional cameras and opens new avenues for research and photography.

\end{abstract}

\maketitle
\section{Introduction}
Over the decades, Embedded Camera Systems~(ECS) have attracted considerable interest~\cite{chen2008citric,cmucam3,rowe2002low,rahimi2005cyclops,wispcam}, leading to their ubiquity. Today,  ECS are used in various applications, including securing of our surroundings~\cite{survcamera,wirelessvid}, documenting lives,  and even equipping insects with vision capability~\cite{insectcam}.  The ECS captures  representations of the physical environment illuminated by light. An array of pixels, forming images, represents the sensed information from ECS. 

Architecturally, an ECS follows a pipelined architecture~\cite{wispcam,rahimi2005cyclops,cmucam3,chen2008citric, naderiparizi2017ultra}, consisting of several stages involving the sensing and processing tasks. One of the critical components in the architecture is the image sensor, consisting of numerous photosensors that track electromagnetic light radiation. Image sensors typically produce weak analog signals, which are then processed by analog front ends. The front-end usually contains a low noise amplifier~(LNA) and automatic gain control~(AGC). The signals are digitized using an analog-to-digital converter~(ADC). The subsequent stage often involves processing the digitized signals using a dedicated chipset to perform tasks such as image compression. Finally, the last steps in the architecture involves further processing of the images, storage, and wireless communication. This ECS architecture uses computational elements, such as a microcontroller or an FPGA, to facilitate several of the steps.

ECS are one of the most widely deployed examples of the wireless embedded system. Nonetheless, we find that the architecture has remained unchanged, presenting challenges.

\fakepar{Challenges} ECS architecture faces limitations, particularly in terms of power consumption. Despite very rapid progress in image sensor technology, where some image sensors now consume few microwatts, the overall power consumption remains  high. This is mainly due to the multiple processing stages required for analog signals generated from image sensors. The processing pipeline typically involves power-intensive components such as LNA, AGC and ADC. An important factor that impacts power consumption is the resolution of the image sensor. A higher-resolution image sensor generates a broader bandwidth analog signal. This requires components with high gain bandwidth product which are energy-expensive. Furthermore, a higher resolution and frequent image capturing escalates the requirement for communication bandwidth. As wireless communication represents the most power-intensive aspect of a WES, and thus transmitting high-resolution images significantly  increases the power consumption of a ECS. 

A second challenge concerns the evolution of camera technology. Over thousands of years, this technology has undergone significant developments, beginning with the camera obscura~~\cite{cameraobscura}, which projected light onto a surface to aid artists, and evolving to the Daguerreotype camera in the early 1800s~\cite{lowry2000silver}, which captured images on silver-coated copper plates. The foundation for contemporary digital cameras was laid by Steven Sasson in 1975 at Kodak~\cite{ElecStillCam}. These advancements, however, have been predominantly focused on capturing visible light (or infrared), thereby covering only a limited aspect of the environment and its various phenomena. Traditional camera technology often overlooks environmental elements such as wind, humidity, temperature, acoustic emissions, magnetic and radio fields, and optical electromagnetic radiations beyond the visible spectrum. This limitation becomes particularly apparent in the advent of extended reality headsets, such as Apple Vision Pro and Xreal Air. Today, these headsets employ conventional cameras to project the physical environment.Nonetheless, this restricts the scope of information displayed on these devices. Unlike human eyes, which are limited to perceiving the narrow range of visible light, these headsets should be capable of visualizing a broader spectrum of our surroundings. Venturing beyond the visible to visualize other phenomena could lead to a more comprehensive understanding of our environment.

\begin{figure*}
    \includegraphics[width=0.8\textwidth]{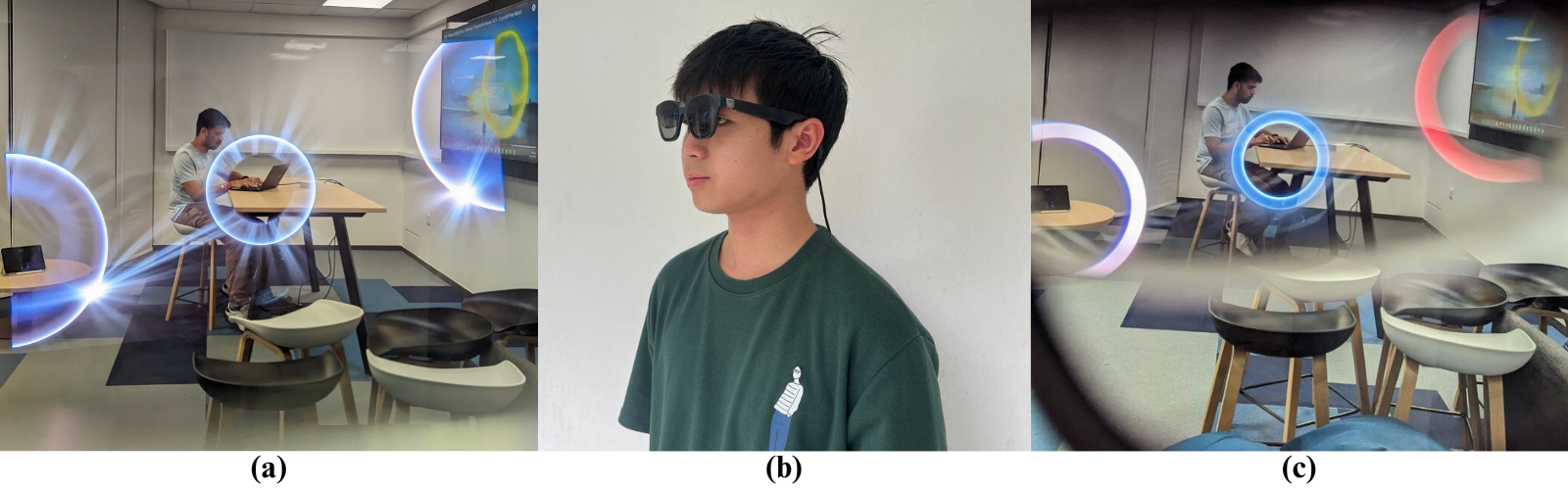}
        \caption{(b) Xreal Air-2 extended reality headset, which we leverage to visualize the acoustic emissions. (a) and (c) exhibit the views from inside the AR glasses, where the user can see the acoustic emission with their eyes. The images are generated using \system. \system\space can facilitate visualization of otherwise invisible sensor streams with applications to extended reality headsets and beyond.}
    \label{fig:banner2}
\subfigure[]{\includegraphics[width=0.3\textwidth]{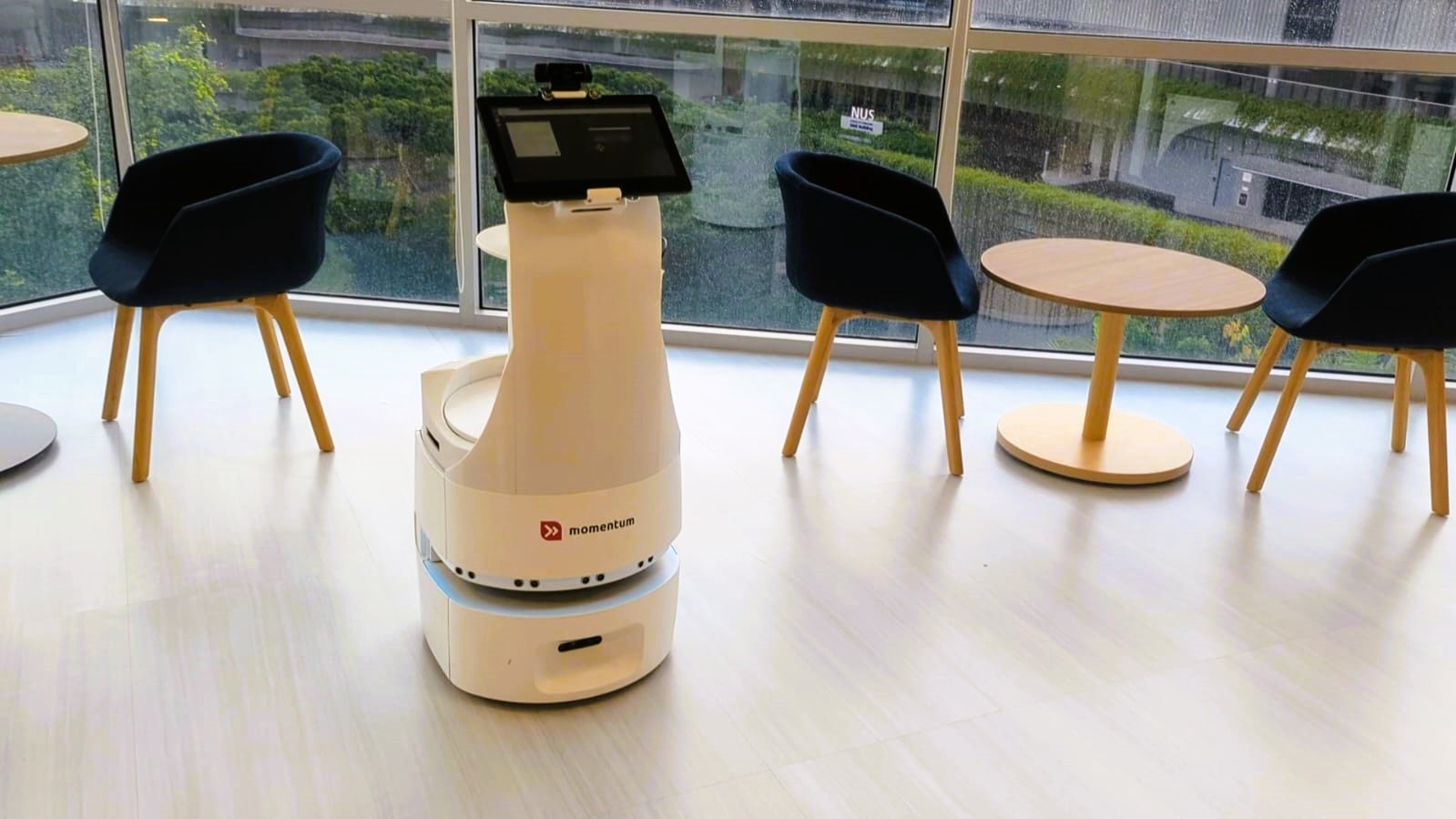}}
\subfigure[]{\includegraphics[width=0.3\textwidth]{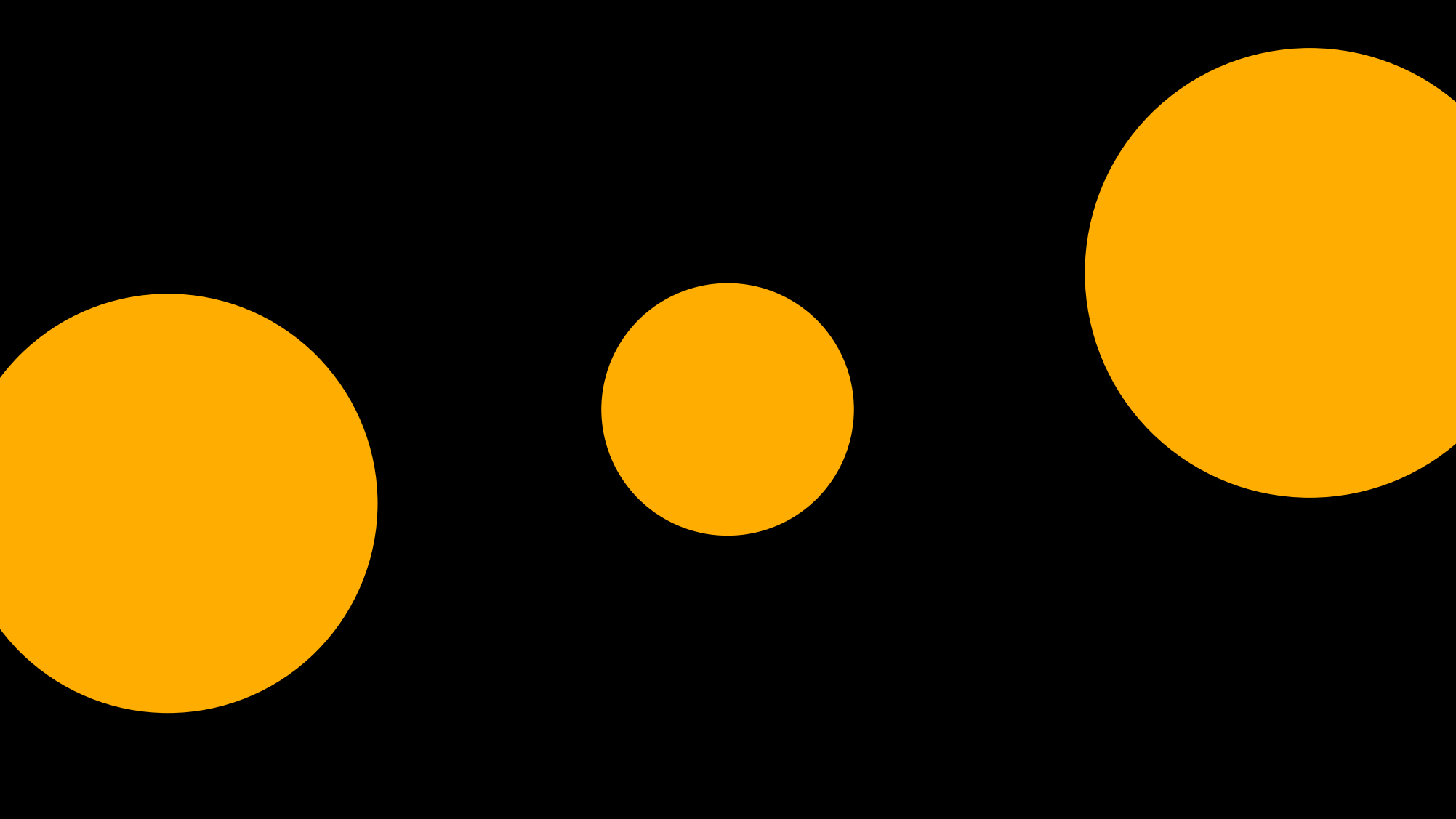}}
\caption{(a) A robot collects information regarding acoustic emissions (b) Acoustic emission extracted and visualised using the \system\space.}
\label{fig:robot_use_case}
\end{figure*}
 
\begin{figure}[t]
\centering     
\subfigure[Conventional architecture]{\label{fig:convention}\includegraphics[width=\linewidth]{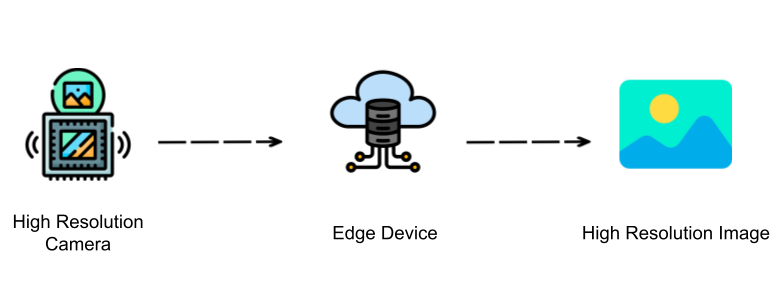}}
\subfigure[\nameF\space architecture]{\label{fig:pixelgen}\includegraphics[width=\linewidth]{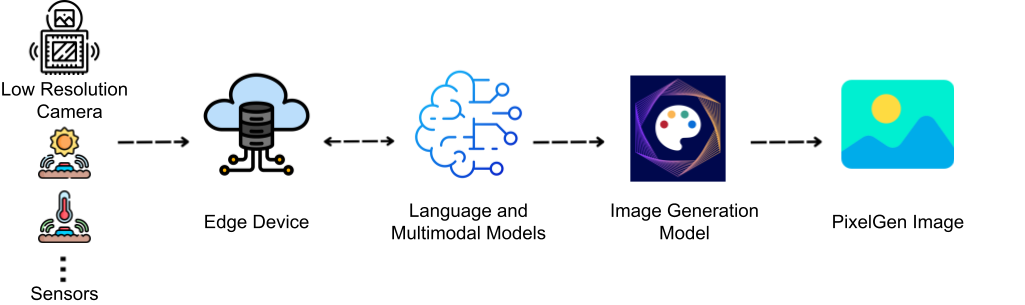}}
\vspace{-4mm}
\caption{\emph{Comparison of ECS architectures with \system}. Conventional architecture capture environmental representations using image sensors that track areas illuminated by visible light. As a result, they do not capture other fields, such as ambient radio waves or vibrations. In contrast, the \system\space architecture employs diverse sensors, including a low-resolution image sensor and radio transceivers, to capture various fields, phenomena, and emissions. Utilizing an LLM, it generates appropriate natural language prompts based on captured data and user input. These prompts are then employed with an image model to generate novel representations of the environment. These representations can also visualize fields, phenomena and emissions that a conventional ECS fail to capture.}
\label{fig:main}
\end{figure}

\fakepar{\system\space design} \system\space represents our take on re-imagining the modern camera. \system\space is a novel ECS architecture that not only facilitates low-power operation but also enables visualization of our surroundings in unprecedented ways.
\system is feasible due to convergence in advances in several fields, including low-power electronics, language and image processing models, and the emergence of mixed-reality headsets. We show the \system\space platform in Figure~\ref{fig:teaser}, and provide compare it against conventional ECS in Figure~\ref{fig:main}.

Language and diffusion models have seen significant interest over the past few years. These advances are built upon the emergence of transformer architecture~\cite{attentiontrans}, which, together with  improvements in compute capabilities and massive amounts of internet based training data, has enabled large transformers based models to demonstrate impressive emergent capabilities. Today, these models enable us to interact with computers using natural language prompts. In another application, they help us identify and understand images. Another class of models, called image models, can generate complex images by taking natural language prompts and images as input. \system\space builds on the emerging capabilities of both language and image models.

At the core of \system\space is the low-power platform , named \hardware, illustrated in Figure~\ref{fig:prototype}.
It is equipped with sensors, transceivers, and low-resolution image and IR sensors, all designed to gather diverse information about the environment. 
This data encompasses a range of fields and phenomena typically overlooked by conventional cameras. 
The selection of sensors and other components for \hardware\space has been strategically done to ensure low power consumption, balancing the trade-off between the bitrate and the capabilities of each sensor. 
For instance, the platform incorporates a low-resolution monochrome sensor. 
This choice is driven by the general principle that power consumption for sensing escalates with the resolution of the image sensor. 
Such design decisions facilitate low power consumption for sensing and communication, as they require transmitting only minimal amounts of data. Wireless communication usually represents the most energy-intensive operation on a WES. 
Therefore, the platform efficiently captures rich environmental information by minimising data transmission needs while maintaining low power consumption.

This leads to the second challenge: representing the collected sensor information. Depending on the application scenarios, the end-user might need to display this information in various ways. In this context, we harness the emerging capabilities of language and image models. We enable the end-user to use natural language to prompt a Large Language Model (LLM), specifying the desired transformations on the sensor data. These transformations could include generating a high-resolution version of an image or representing acoustic emissions.  The transformed data and instructions are then fed to an image model as prompts. This process allows the model to generate unique  representations. Our paper demonstrates that this approach can produce high-resolution images from low-resolution monochrome inputs and other environmental data or visualize acoustic emissions.

\fakepar{Summary of results} We provide a summary of important results presented in this work:
\begin{itemize*}

\item \system\space is the first embedded platform designed to integrate with image and language models, enabling the generation of unique representations of the environment.
\item \system\space facilitates the generation of high-resolution images using only low-resolution,  monochrome images and basic environmental information captured by the platform.
\item \system\space enables the visualization of phenomena typically invisible to eye, such as acoustic emissions.
\end{itemize*}

\fakepar{Application use case} \system\space can facilitate various application scenarios. These range from monitoring the environment in a privacy-preserving manner through robots to designing cameras capable of streaming high-resolution images while operating for extended periods on a battery. Nonetheless, in this work, we prototype an application that visualizes invisible environmental fields, such as acoustic emissions, through a mixed-reality headset. We show these  application scenarios in Figure~\ref{fig:banner2} \& ~\ref{fig:robot_use_case}.

\section{Related work}

We discuss the works closely related to \system, and also place \system\space in context to them.

\fakepar{Embedded camera systems} ECSs are ubiquitous and play a critical role in ensuring the safety and monitoring of significant assets. Research and development in ECS over the past two decades have made this possible. Cyclops~\cite{rahimi2005cyclops} was an early design of an ECS that coupled an image sensor with a CPLD for image processing, thus easing the processing burden on the microcontroller. CITRIC coupled an image sensor to a powerful microcontroller, large storage and 802.15.4-based transceiver to enable the design of a camera-based sensor network~\cite{chen2008citric}. In another effort, CMUCam was an open-source ECS optimised for the processing of images, including supporting capability for near real-time processing~\cite{rowe2002low, cmucam3}. Nonetheless, these platforms were significantly power consuming, with average power consumption in tens to hundreds of milliwatts. Furthermore, they only captured information from a visible light-based image sensor. \system\space builds on these efforts; however, it proposes an entirely new architecture of ECS that captures information through an array of sensors and uses an image model to generate images. As we demonstrate in this work, this allows us to lower the power consumption and bandwidth constraints and to also represent phenomena that are not visible to an image sensor. As an example, to visualise acoustic emissions.

\fakepar{Low-power ECS} Recent efforts have focused on reducing the power consumption of ECS. The WISPCam~\cite{wispcam}, based on an RFID system, is an example of a battery-free camera that harvests radio energy and stores it in a capacitor. It communicates captured images back to the RFID reader using the EPCGen protocol. Naderiparizi et al. designed a camera capable of streaming high-definition images with minimal power consumption~\cite{naderiparizi2017ultra}.  Using a backscatter mechanism they eliminate computational blocks from the camera architecture. This camera, however, has a limited communication range of only a few meters. Using the LoRa protocol, Camaroptera~\cite{camptoreia} represents another approach, transmitting low-resolution images over long distances with power harvested from a solar cell. By heavy-duty cycling and by taking infrequent, low-resolutions images, Camaroptera achieved battery-free operation.

However, these systems have limitations. The scope of their camera is primarily limited to the visible spectrum. To conserve power, they sacrifice image resolution or duty cycle. While Naderiparizi et al.'s camera can capture high-resolution images~\cite{naderiparizi2017ultra}, its backscatter mechanism limits its  range to a few meters. In contrast, \system\space overcomes these constraints by capturing a broader  representation through a low-resolution image sensor, environment and other sensors, and radio transceivers. It leverages the capabilities of an image model to produce higher-resolution images from this data. As \system\space does not capture high-resolution images, or require high-bandwidth for communication, it can support low-power operations without  sacrificing of the image quality for the end-user.


\fakepar{Other systems} In 2013, LiKamWa introduced an innovative approach to image processing to enhance energy efficiency in camera systems. This technique involved resizing images during inactivity, specifically when no ambient motion was detected. Building on this concept,  Naderiparizi et al. proposed "Glimpse"~\cite{naderiparizi2017glimpse}, a system that employs an "early discard" architecture. In Glimpse, the ECS automatically discard an image if it detects no significant changes in lighting conditions or the absence of ambient motion. Ila leverages image stacking and batching to combine several individual frames into one frame (similar to a mosaic) to expedite the processing pipeline~\cite{gokarn2023mosaic}. Jiyan proposed MRIM, an exciting framework incorporating software implementation of the early discard locally at the ECS and only transmits informative frames to the edge for additional processing~\cite{wu2022mrim}. Some also work on image enhancement techniques like super-resolution and leveraging dual camera setup to get a stream of low-resolution frames with intermediate high-resolution frames \cite{veluri2022neuricam}.

\section{Design}
\system\space features a custom hardware platform called \hardware\space and an edge computer. The \hardware\space platform captures various environmental data, including low-resolution images, temperature, humidity, light intensity, and radio, magnetic, and acoustic emissions. This data is transmitted to the edge device in an energy-efficient manner. Using natural language, the user prompts the language model on the edge device to transform the sensor data. The language model generates prompts for a diffusion model. The prompt and the sensor data enable the diffusion model to create visual representations of the environment. We show an overview of the system in the Figure~\ref{fig:pixelgen}.

\subsection{\hardware}
We describe the design and implementation of \hardware, and place \hardware\space in context to conventional ECS.

\fakepar{Overview of ECS architecture} A conventional ECS architecture follows a pipelined architecture. The process begins with an image sensor capturing light intensity values, which are then amplified and stabilized through AGC. These light readings are digitized, allowing for image compression or other processing to be conducted. A dedicated chipset often performs these steps. The microcontroller manages the processed image for storage and wireless transmission. Even though the image sensor consumes relatively little power, hundreds of microwatts for a high-resolution sensor are power-consuming, and subsequent processing steps, especially wireless transmissions. Consequently, these systems often require external power sources or large batteries, also affecting their overall form factor. Additionally, the ECS typically captures only a narrow spectrum of the environment, limited to what is illuminated by visible or infrared light. Typically, there has been trend of increasing the pixel count for the ECS, and hence the power consumption. We  illustrate the architecture of a conventional ECS in the Figure~\ref{fig:convention}.

Recent works focus on lowering power consumption, mainly through integrating a backscatter mechanism within the ECS architecture. The backscatter mechanism enables transmitting data at microwatts of power consumption by reflecting or absorbing the ambient wireless signals~\cite{loreaback,ambientback}. One promising approach involves the direct interfacing of the image sensor with the backscatter module, eliminating the computational elements~\cite{naderiparizi2017ultra,bfcellphone,bfvls}. 

\begin{figure}[t]
    \centering
    \includegraphics[width=0.6\columnwidth]{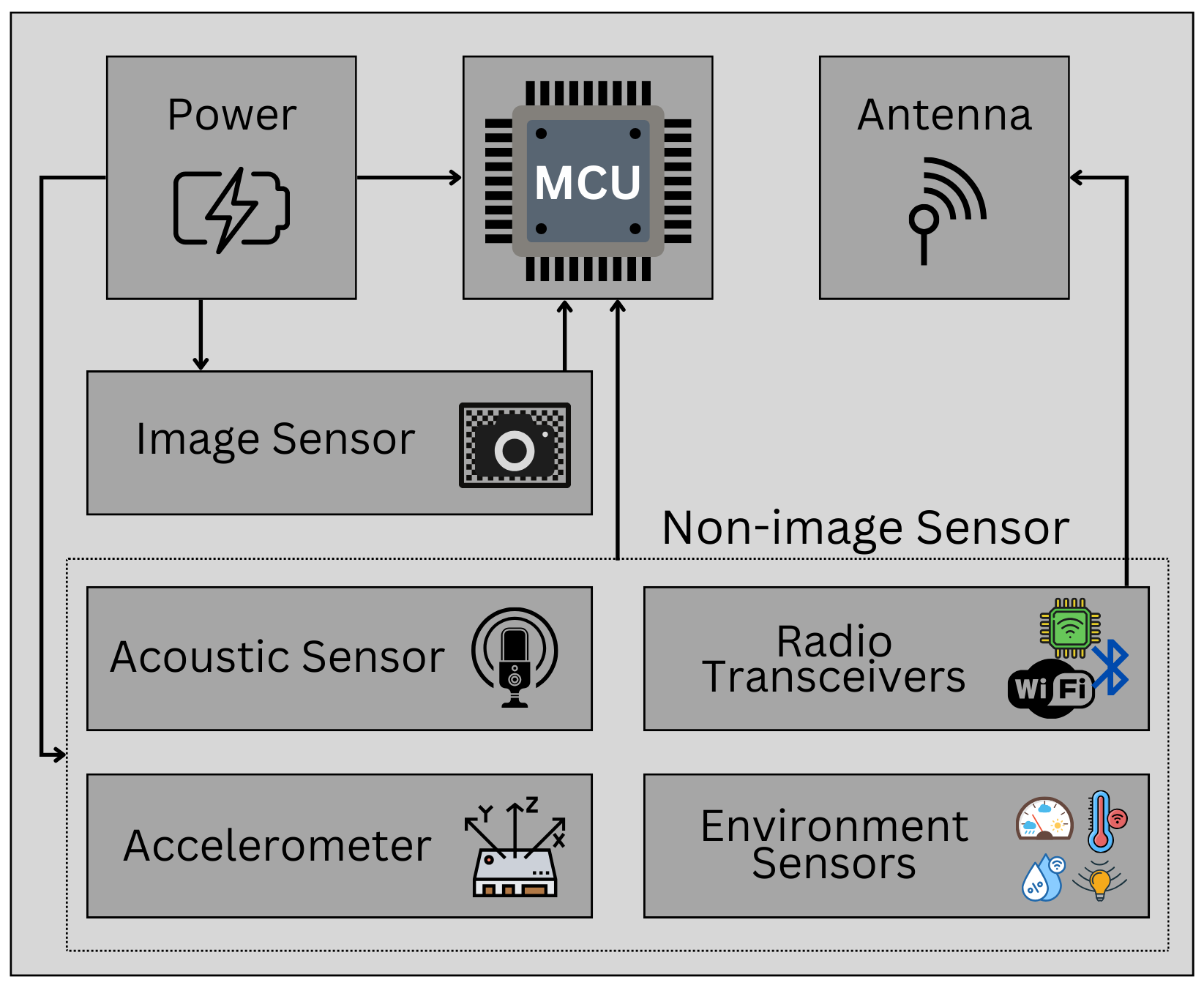}
    \caption{\hardware\space  is composed of  sensors, microcontroller and transceivers, possibly supplemented with image sensors. Its primary function is to gather diverse environmental data. The system is designed for energy efficiency, potentially using a backscatter mechanism to enable low-power communication and operation on harvested energy. A key feature of \hardware\space is its use of a diffusion model to reconstruct images from the collected sensor data. Unlike conventional embedded cameras, \hardware\space may not even sport a image sensor on the platform.}
    \label{fig:blk_diagram_pixelgen}
    \vspace{-4mm}
\end{figure}

\fakepar{Design of platform}  \hardware\space borrows features from conventional and backscatter-based ECS architecture. \hardware\space can utilize a low-power image sensor to capture physical environment representations in the visible spectrum and may employ a backscatter mechanism for energy-efficient images and sensor data communication. However, \hardware\space goes beyond these ECS architecture significantly. It captures a  comprehensive representation of the environment. The platform can include various sensors: transceivers for radio and magnetic emission tracking, microphones for acoustic emissions, ambient sensors for physical environmental parameters, and a low-resolution image sensor. This diverse sensor array allows \hardware\space to capture a broader range of environmental data than standard cameras, which typically focus only on the visible spectrum. Specific sensors can be selected based on the application's requirements and constraints regarding bandwidth and power consumption. For instance, the system might only capture low-resolution images with ambient light but can enhance these images using cues from other sensors through a language and image model. Hardware design and bandwidth constraints are greatly simplified by communicating only low-resolution images with simple sensor readings. We show a high-level schematic of the \hardware\space platform in the Figure~\ref{fig:blk_diagram_pixelgen}.


\begin{figure}
\centering     
\includegraphics[width=0.48\textwidth]{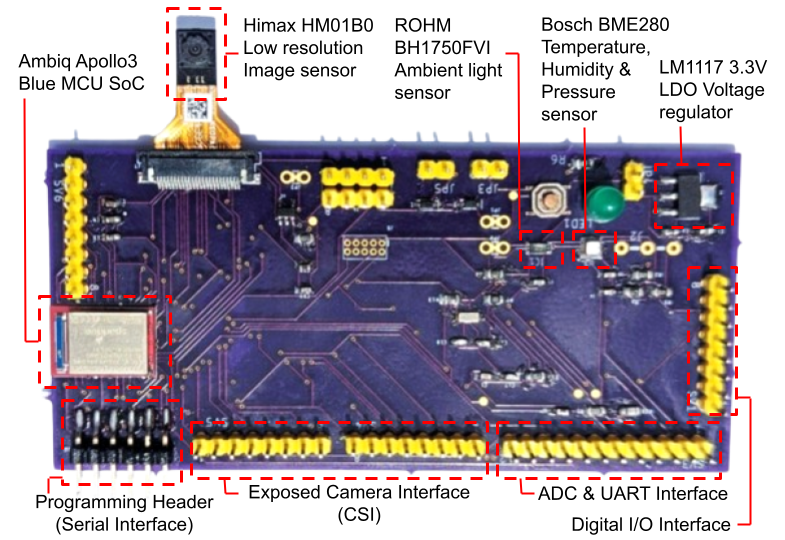}
\vspace{-4mm}
\caption{\hardware\space platform - Custom board that enable PixelGen to generate an image from a low-power camera and an array of sensor data.}
\vspace{-4mm}
\label{fig:prototype}
\end{figure}

\fakepar{Implementation}  
\hardware\space hardware is designed and tailored for specific scenarios. 
It enables the collection of vision information and some onboard processing. The hardware is optimised to ensure prolonged operation on a small battery. We employ a powerful yet highly energy-efficient Ambiq Apollo3 blue microcontroller~\cite{ambiq2023}. We equip the platform with Bosch BME280 \cite{bosch2023} to measure temperature, humidity, \& pressure information from the environment. We estimate the ambient light conditions using a ROHM BH1750 sensor~\cite{rohm2023}. In addition, it can monitor acceleration using the STM LSM9DS1 9-axis IMU sensor. The platform also has a low-resolution monochrome image sensor, Himax HM01B0 \cite{himax2023}. The platform can also perform acoustic emissions sensing. From the perspective of wireless communication, the platform can leverage either the backscatter mechanism or the Bluetooth transceiver part of the Apollo3 microcontroller. 

We have designed the platforms on an FR4, 4-layer PCB manufactured by OSHPark \cite{oshpark2023}. We show this in Figure~\ref{fig:prototype}. 


\begin{figure}[t]
    \centering
    \includegraphics[width=0.9\linewidth]{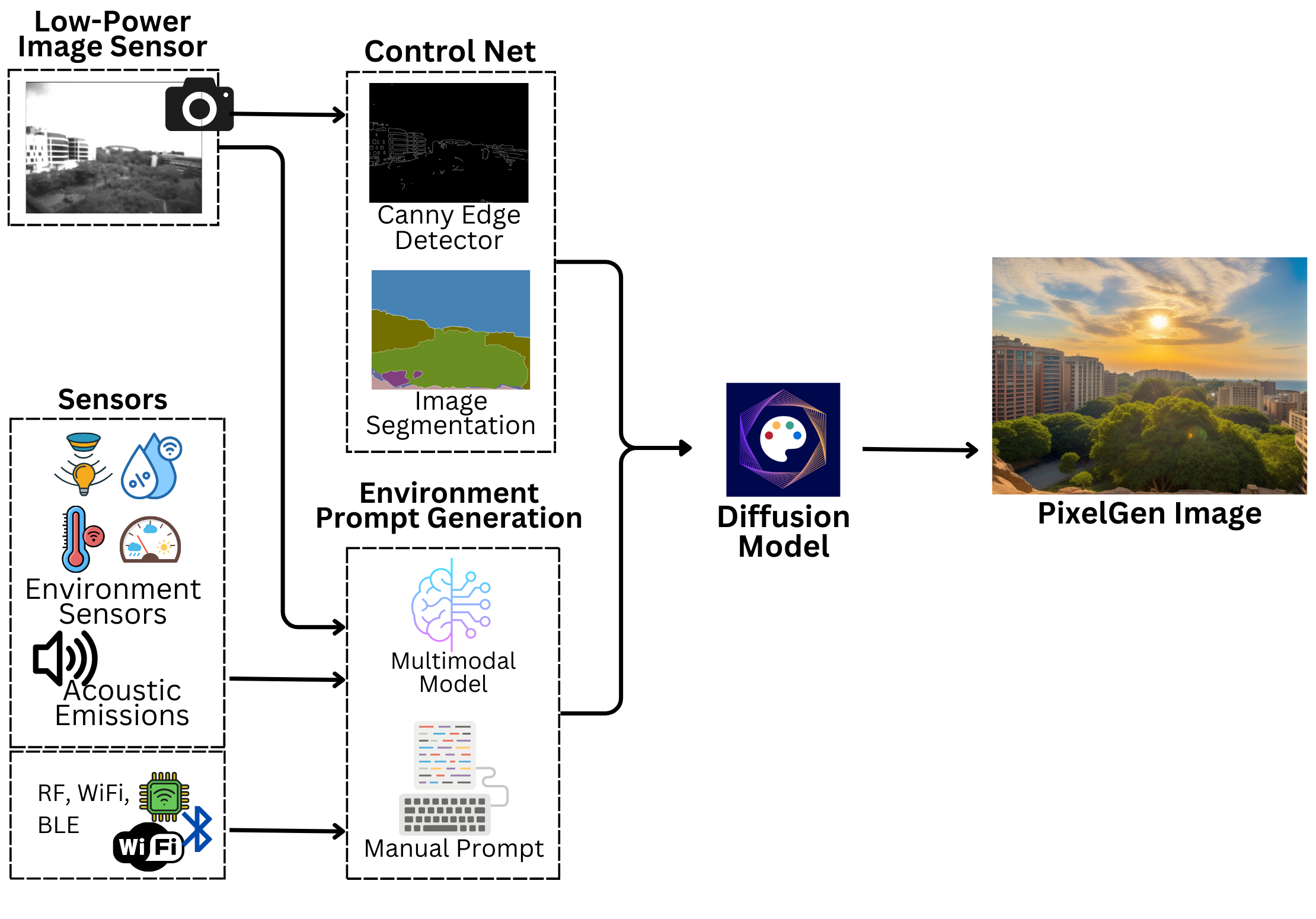}
    \caption{The edge computer facilitates interaction between the end-user and the \system. It receives inputs from the end-user through natural language prompts, providing instructions on utilizing the collected sensor data. These prompts are paired with sensor data and fed into a language model, which generates specific prompts for the diffusion model. Using the image, sensor data, and generated prompts, the diffusion model processes this information to create a rich representation of the physical environment. As an example, the end-user may want to provide prompts to generate a high-resolution image even though \hardware\space only captured low-resolution image.}
    \label{fig:edge_computer_architecture}
\end{figure}

\subsection{Edge Computer}

\begin{table*}[ht!]
\begin{tabular}{|c|c|c|c|c|}
\hline
Brand Name         & Model Number \& Specifications                                                                        & Processor Clock Speed & Memory & \begin{tabular}[c]{@{}c@{}}Generation Time\\ (seconds)\end{tabular} \\ \hline
Apple MAC Mini     & \begin{tabular}[c]{@{}c@{}}M2 Pro with 10-core CPU, 16-core GPU,\\ 16-core Neural Engine\end{tabular} & 3.49 GHz              & 32 GB  & 90                                                                  \\ \hline
Legion Y9000X IAH7 & \begin{tabular}[c]{@{}c@{}}Intel i7-12700H Processor with Nvidia \\ GeForce RTX 3060\end{tabular}     & 4.70 GHz              & 16 GB  & 31                                                                  \\ \hline
\end{tabular}
\caption{Different mini PC configurations for implementing edge device. We denote the time for generating images using image model.}
\label{tab:minipc}
\end{table*}

The edge computer receives information from \hardware\space platform and also interacts with the end-user.

\fakepar{Overview} The \hardware\space platform transmits sensor information to the edge computer, serving as a repository for aggregating this data. Users interact with the system by providing natural language prompts to a language model operating on the edge device. These prompts guide the language model to create specific instructions for processing the collected sensor data. Combining these user-directed prompts with sensor data forms an input for a image model. This model processes the combined input to generate detailed images of the environment. For example, an input from the user might direct the language model to generate instructions for the diffusion model to transform a low-resolution image into a high-resolution. The system effectively leverages the strengths of language and image models, using user input to bridge the gap between raw sensor data and the desired  representation of the environment. The edge computer's steps and architecture are depicted in Figure~\ref{fig:edge_computer_architecture}.

\fakepar{Using language models to generate prompts}  The aggregated sensor data, such as those from environmental sensors or acoustic emissions, need to be converted into suitable prompts for the image model. For this purpose, we utilize an LLM at the edge computer. The LLM can either run locally or be accessed remotely through API calls. Using an LLM serves distinct purposes: 1) It lowers the entry barrier for the end-user, who does not need to learn about the system intricacies and can provide instructions in natural language to process image and sensor data. 2) LLMs can be trained and fine-tuned to align with the image model's specifics, ensuring reliable image generation. 3) Multimodal LLMs can analyze the collected images, thus generating specific prompts.

\fakepar{Image model} To visualize the surroundings, we employ an image model. This model is trained on a vast collection of images and can be prompted using natural language, images, or a combination of both to generate unique images. We utilize the image model in \system\space for image generation. Specifically, the model receives sensor and image data and a suitable prompt generated by the LLM as inputs. Utilizing these inputs, the model creates representations of the world. This process happens in two discrete steps. 

Nonetheless, In the first step, we process the low-resolution monochrome images. We use ControlNet to retrieve additional information from the low-resolution  images. ControlNet is a neural network framework~\cite{zhang2023adding} that controls diffusion models by adding extra conditions. This includes extracting edges, boundaries, objects and other such information. The ControlNet can be used with different preprocessing models, and we demonstrate it in the evaluation section. For instance, a Canny edge detector can help retain the composition of an image by extracting the outline and edges. Likewise, the segmentation preprocessor OneFormer can transfer the location and shape of objects by labeling them with different predefined colors. 

In the second step, we integrate prompts derived from environmental sensor data and augment this with additional control from the original low-resolution monochrome images. Using the diffusion model, this approach enables us to produce high-resolution, multimodal RGB images. While maintaining the original composition details of objects and backgrounds, we enhance these images with added colour information and improved resolution.

\fakepar{Implementation} We implement the language model using the OpenAI GPT. Nonetheless, for generality, we also implemented a LLM based on Llama. As an image model, we use the stable diffusion~\cite{stablediffusion2023,stablediffusionwiki}, which is an open-source diffusion model with weights Realistic Vision V5.1 \cite{civitai2023}.  As an edge device, we selected mini computers that are capable of operating the stable diffusion model locally. In particular, we explore two mini computers as shown in the table~\ref{tab:minipc} with the corresponding execution time for stable diffusion.



\section{Evaluation}
We evaluate \system\space in various scenarios. 

\begin{figure*}[ht]
\centering     
\subfigure[]{\includegraphics[width=0.23\textwidth]{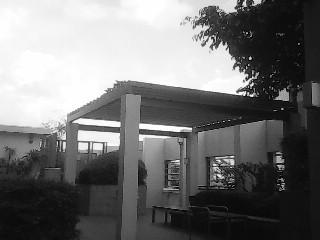}}
\subfigure[]{\includegraphics[width=0.23\textwidth]{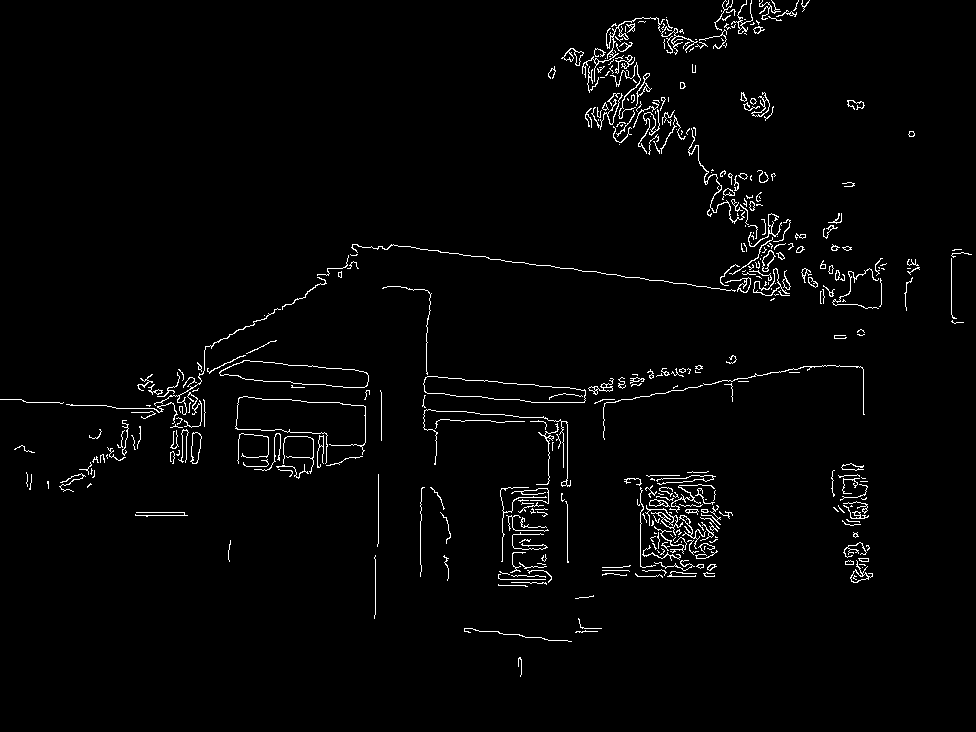}}
\subfigure[]{\includegraphics[width=0.23\textwidth]{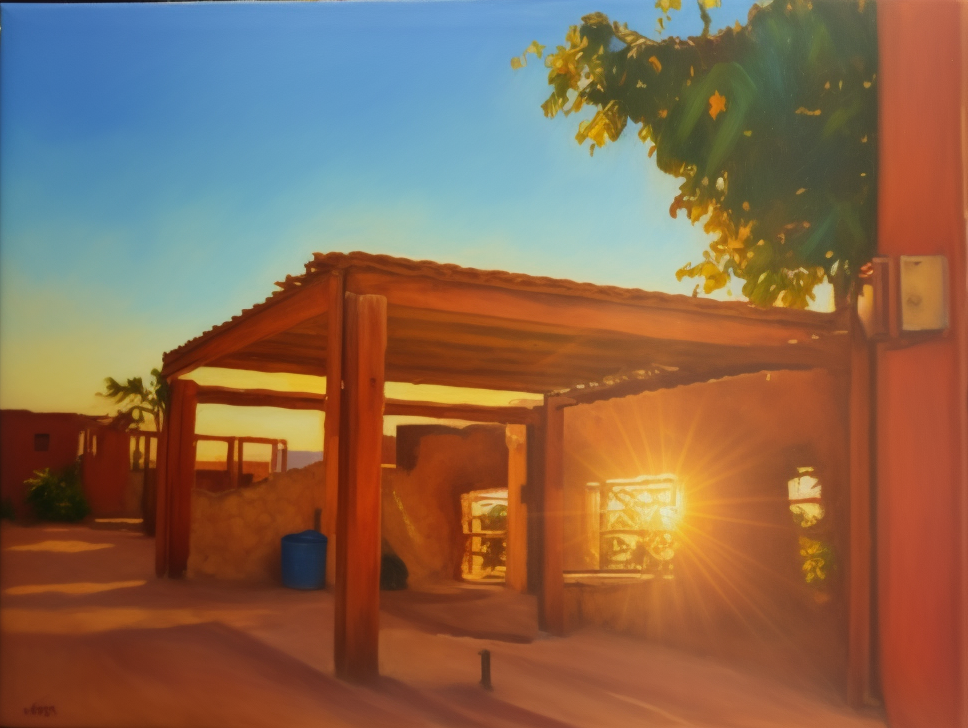}}
\caption{(a) shows the original monochrome image (324x244) provided as an input to the canny edge detector (ControlNet), which extracts the composition of the image in terms of edges, shown in (b) which we combine with environmental sensor stream (Ambient Light: 8407 Lux, Temperature: 29.52 $^{\circ}$C; Humidity: 63.11\%, Pressure: 1006.87 hPa, and Wind Velocity: 0.0 $m/s$) as a manual prompt to the stable diffusion model to generate (c) an artistic image (968x728) as an output.}
\label{fig:E1}
\end{figure*}




\begin{figure*}[h]
\centering     
\subfigure[]{\includegraphics[width=0.23\textwidth]{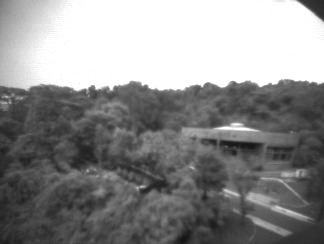}}
\subfigure[]{\includegraphics[width=0.23\textwidth]{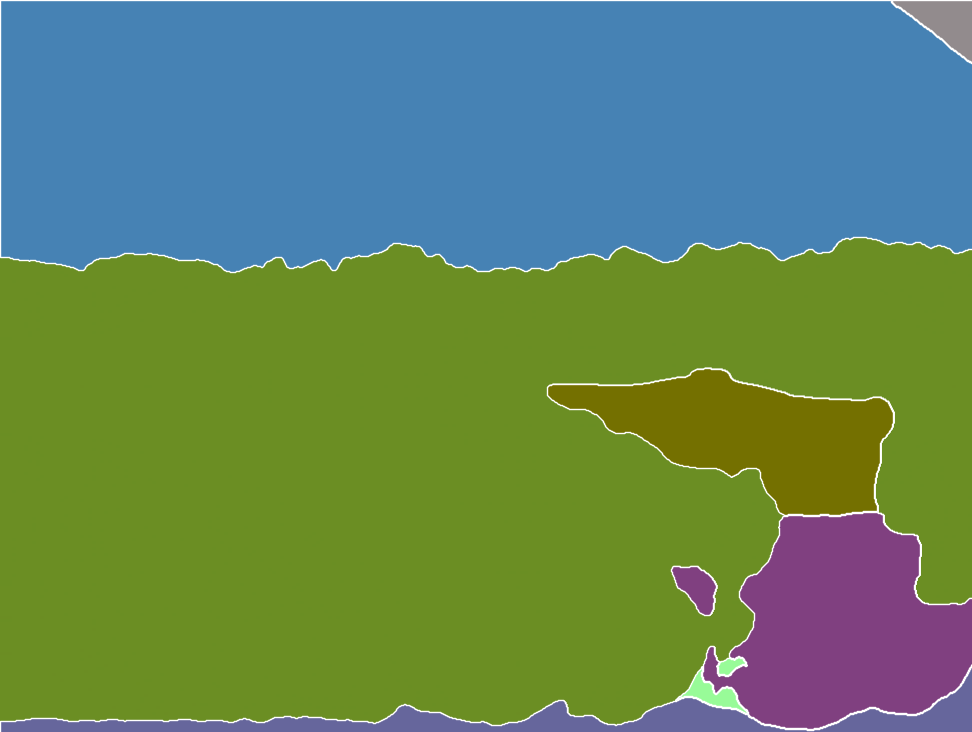}}
\subfigure[]{\includegraphics[width=0.23\textwidth]{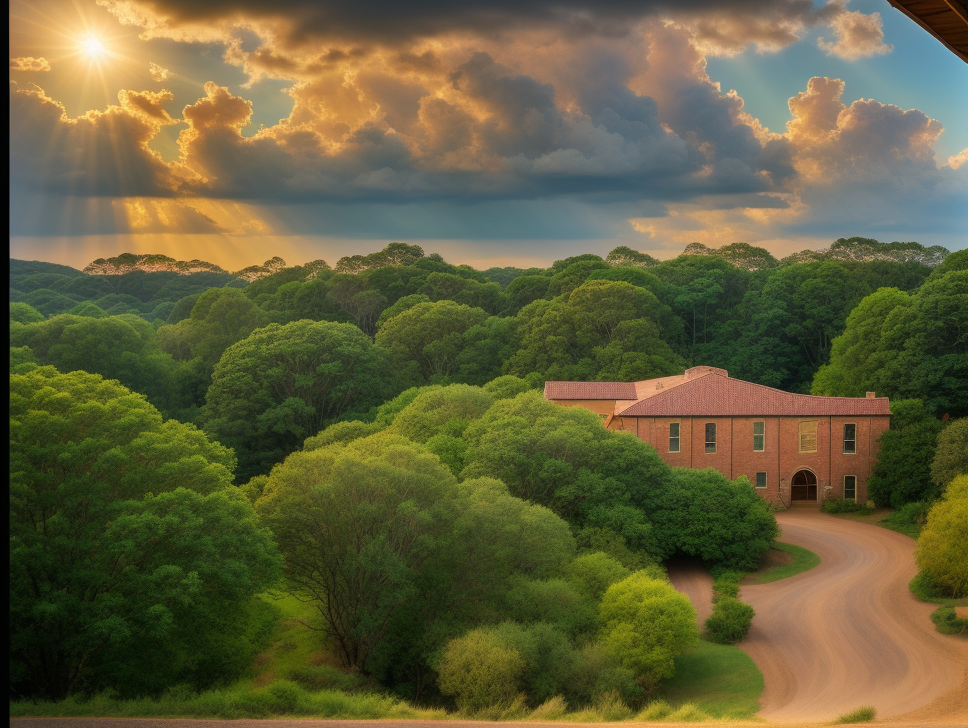}}
\caption{(a) shows the original monochrome image (324x244) provided as an input to the OneFormer (Coco) Segmentation Model (ControlNet), which extracts the location and shape of the objects, shown in (b) which we combine with environmental sensor stream (Ambient Light: 2372 Lux, Temperature: 29.82 $^{\circ}C$; Humidity: 66.52\%, Pressure: 1002.98 hPa, and Wind Velocity: 2.2 $m/s$) as an automated prompt generated by GPT-4 to the stable diffusion model to generate (c) a realistic image (968x728) as an output.}
\label{fig:E2}
\end{figure*}

\fakepar{Setup} We used  pre-processing models to extract information from the captured grayscale image. For the experiments discussed in this section, we employed Canny Edge Detection Model~(CEDM), Oneformer Segmentation Model~(COCO), and the Line Art Control Model~(LACM). These  models are tasked with extracting meaningful information about the objects in captured grayscale images, including their position, edges, colour palettes, and other pertinent features such as brightness, hues, and darkness. In the next step, the captured information from sensors is converted to prompt with the involvement of a language model or manually. In the final step, the prompt and other information are provided to the image model, i.e., stable diffusion. This leads to the generation of high-resolution images that represent the environment.  In particular, we have used the variant of stable diffusion, called Realistic Vision 5.1. This model allows generation of images in various styles, and its behaviour is controlled using various negative/positive prompts. Nonetheless, the design of \system\space is generic, and can use other image models. 

We evaluated the ability of \system\space to generate diverse and novel representations. Hence, we narrowed it down to four styles: artistic, realistic, traditional Chinese painting, and realistic Van Gogh oil painting. We collected 126 monochrome images and sensor data~(temperature, humidity, pressure, and light intensity) in different locations and diverse environmental conditions. The original images are 324 x 244 in resolution and have a monochrome format. Post-processing, \system, regenerates high-resolution images with different resolutions. For brevity, we only illustrate results obtained at the resolution of 968 x 728, i.e., approximately three times higher pixel count of the captured image.

\fakepar{Generating realistic images} We investigate \system\space ability to generate high-resolution and realistic visualisation of the environment. We capture monochrome images and environmental data from the \hardware\space. We configure the image model to generate realistic images. We preprocess the captured monochrome image with the OneFormer~(COCO) Segmentation model. This model extracts the location and shape of objects in the image. We collate this information with an appropriate prompt generated using an LLM and environmental sensor data and give it as an input to the image model. Figure~\ref{fig:E2}(a) shows the original input image, and Figure~\ref{fig:E2}(b) shows the segmented output of the COCO model, along with \ref{fig:E2}(c) the generated image from the image model.

\fakepar{Generating artistic images} We assess \system\space capability to visualize the environment in an artistic style. We first acquire a monochrome image using \hardware, followed by preprocessing it with a CEDM. The extracted information from this process is then combined with data from environmental sensors and a manually provided prompt, serving as an instruction to the image model. We show the original image, the output from the CEDM, and the subsequently generated images in Figure~\ref{fig:E1}. Notably, the captured image is in monochrome, and the additional input from the environmental sensors furnishes the image model with context. This enables \system\space to generate a higher-resolution and artistically rendered version of the surrounding environment.

\begin{figure*}[h]
\centering     
\subfigure[]{\includegraphics[width=0.23\textwidth]{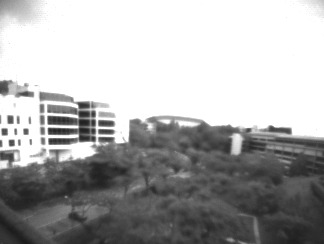}}
\subfigure[]{\includegraphics[width=0.23\textwidth]{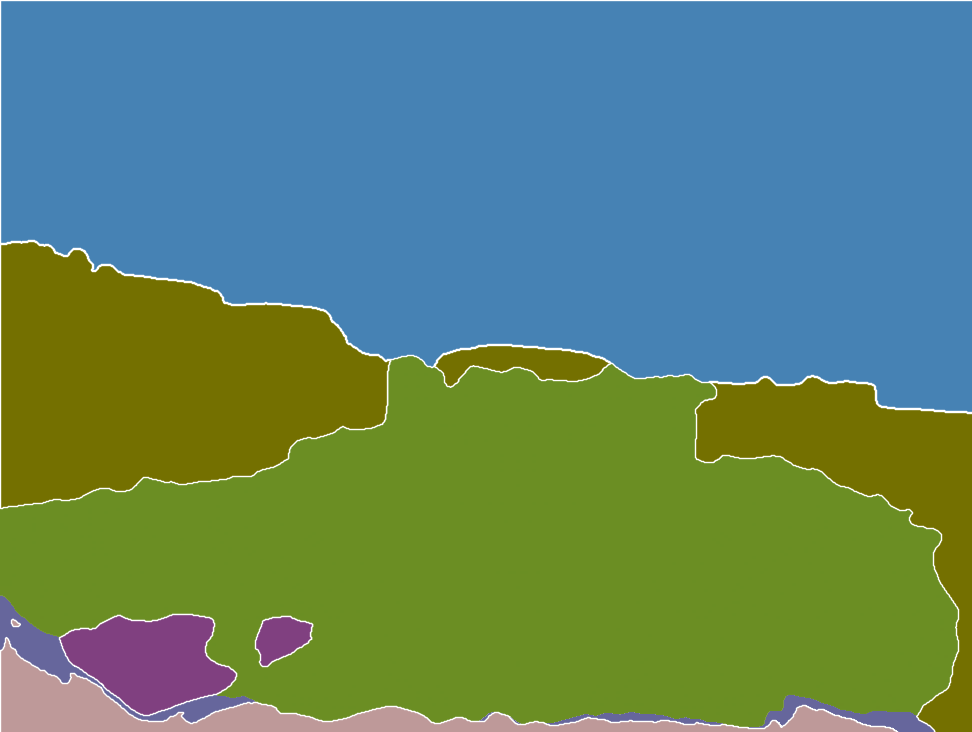}}
\subfigure[]{\includegraphics[width=0.23\textwidth]{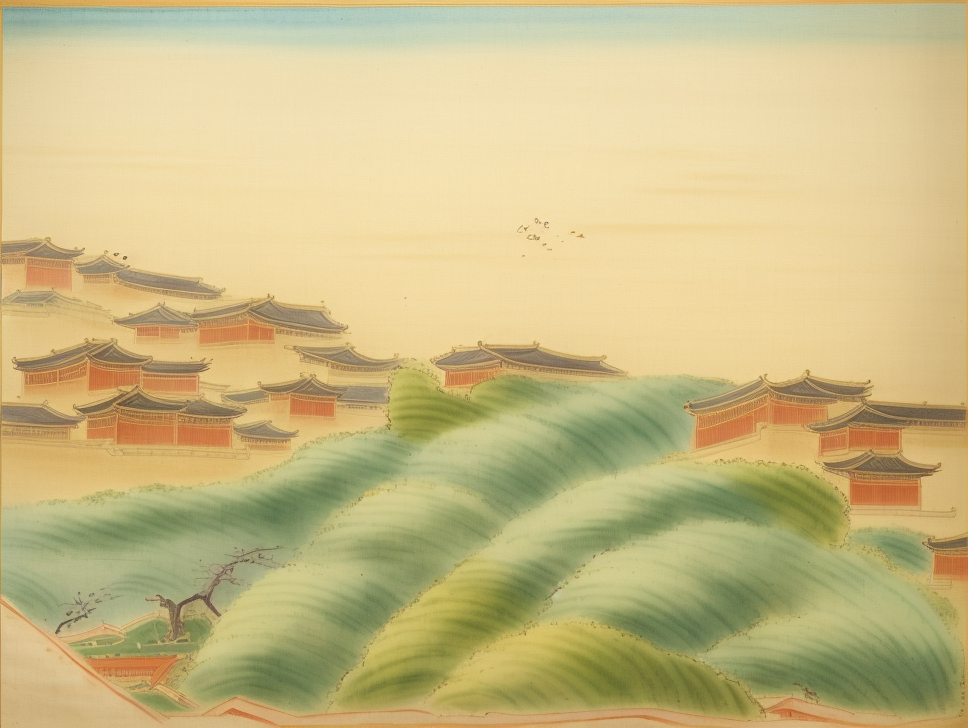}}
\caption{(a) shows the original monochrome image (324x244) provided as an input to the OneFormer (Coco) Segmentation Model (ControlNet), which extracts the location and shape of the objects, shown in (b) which we combine with environmental sensor stream (Ambient Light: 2372 Lux, Temperature: 29.82 $^{\circ}C$; Humidity: 66.52\%, Pressure: 1002.98 hPa, and Wind Velocity: 2.2 $m/s$) as an automated prompt generated by GPT-4 to the stable diffusion model to generate (c) a Traditional Chinese Painting (968x728) as an output.}
\label{fig:E3}
\end{figure*}

\fakepar{Generating Chinese painting styled images} We explored the ability to generate traditional Chinese painting-styled images. We keep the sensor data, image, and model~(COCO) similar to the earlier experiment. Figure~\ref{fig:E3}(a) shows the original image, and ~\ref{fig:E3}(b) shows the segmented output from COCO model, similar to ~\ref{fig:E2}(b), and the resulting image in the  Figure~\ref{fig:E3}(c). We observe that the rendered painting has an identical vegetation placement to the  monochrome image. We also demonstrate an actual prompt that we leverage to generate the traditional Chinese art in Figure~\ref{box}.
%
\begin{tcolorbox}[colback=white,colframe=black,arc=0pt,outer arc=0pt,boxrule=0.5pt,title=Sensor Value,fontupper=\ttfamily,breakable]
Ambient Light: 2372 Lux;\\
Temperature: 29.82 $^{\circ}C$;\\
Humidity: 66.52 \%;\\
Pressure: 1002.98 hPa;\\
Wind Velocity: 2.2 $m/s$;\\ \\
\textbf{Prompt from GPT-4:} \\ \\
"A traditional Chinese painting depicting a warm, slightly humid day with a gentle breeze. The scene, illuminated by a moderate level of light, features a harmonious blend of natural elements. The artwork uses delicate brushwork and ink washes to capture the movement of air and the subtle warmth, set in a serene landscape typical of traditional Chinese art."
\end{tcolorbox}
\noindent
\begin{minipage}{\linewidth}
\captionof{figure}{An actual prompt that we leverage to generate the traditional Chinese art.}
\label{box}
\end{minipage}

%
%

\begin{figure*}[h]
\centering     
\subfigure[]{\includegraphics[width=0.23\textwidth]{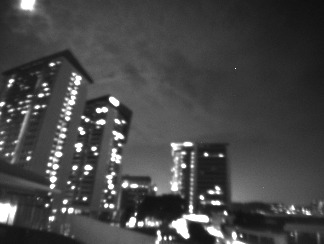}}
\subfigure[]{\includegraphics[width=0.23\textwidth]{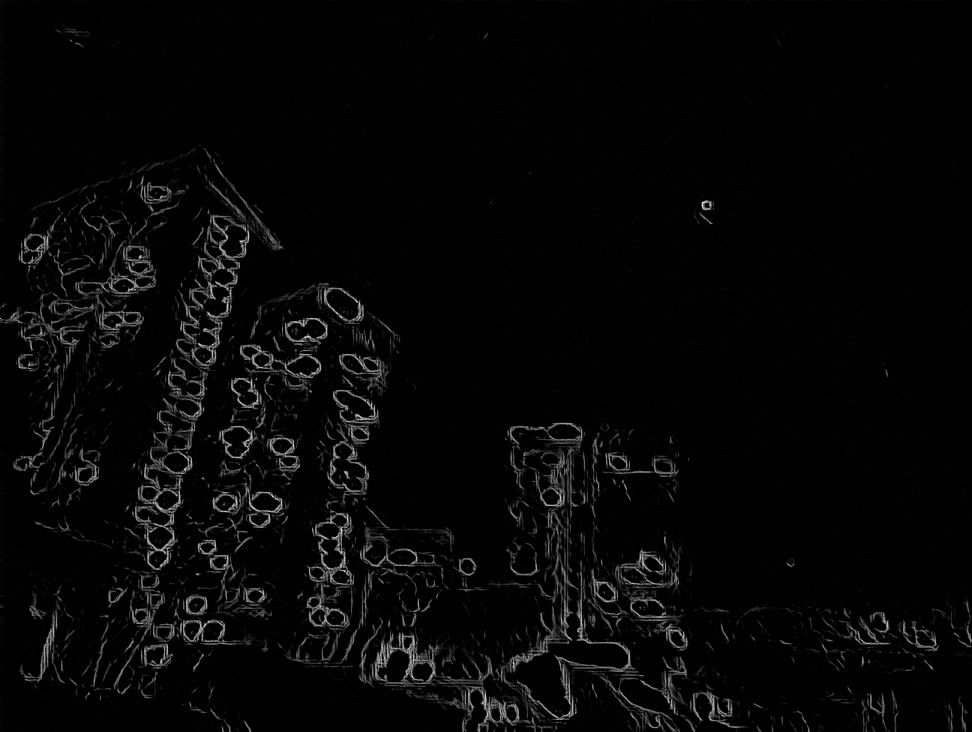}}
\subfigure[]{\includegraphics[width=0.23\textwidth]{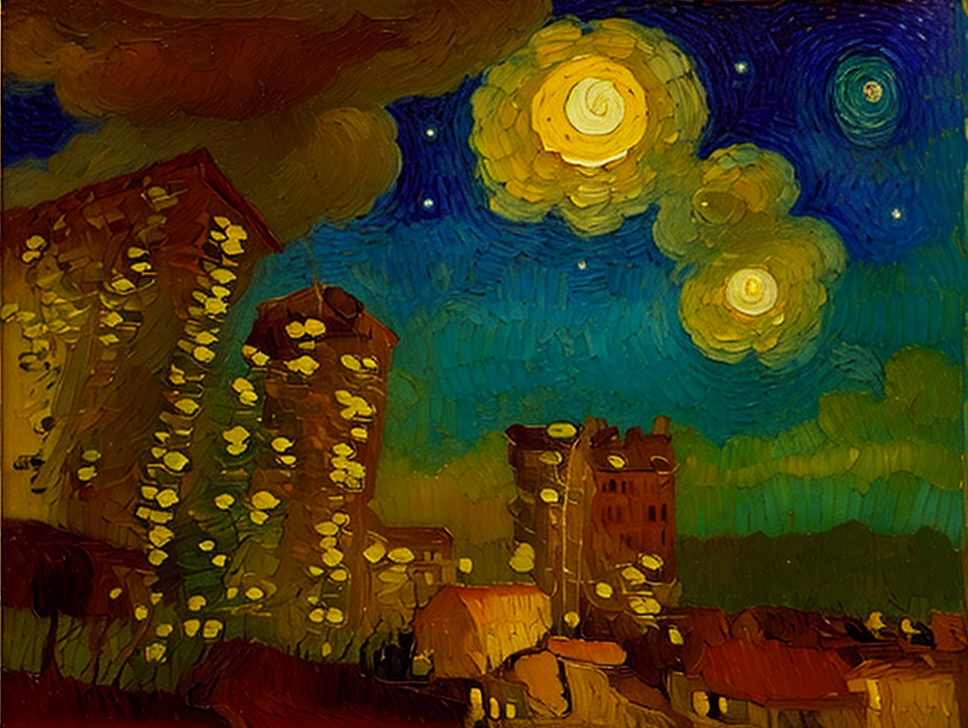}}
\caption{(a) shows the original monochrome image (324x244) provided as an input to the Line Art Control Model (ControlNet), which extracts essence of the still image like hues and dark along with the location and shape of the objects, shown in (b) which we combine with environmental sensor stream (Ambient Light: 0 Lux, Temperature: 28.51 $^{\circ}C$; Humidity: 71.37\%, Pressure: 1008.75 hPa, and Wind Velocity: 0.0 $m/s$) as an automated prompt generated by GPT-4 to the stable diffusion model to generate (c) a realistic Van Gogh Oil Painting (968x728) as an output.}
\label{fig:E4}
\end{figure*}
\fakepar{Generating Van-Gogh inspired oil painting} This is the final experiment to investigate the ability of \system\space generate artistic images. In this particular experiment, we extract the image-related features using the LACM. We feed the extracted feature using a prompt generated using a LLM, and the corresponding environment data. We show the results of the experiment in the Figure~\ref{fig:E4}. The Figure~\ref{fig:E4}(a) shows the captured image, Figure~\ref{fig:E4}(b) shows the output of the LACM, and the Figure~\ref{fig:E4}(c)  demonstrates the output of the image model. We notice that the features extracted by LACM highlights the essence of the still image like hues and dark along with the location and shape of the objects. The generated image is in the style of an oil painting.\\
\begin{figure*}[ht!]
\centering     
\subfigure[]{\includegraphics[width=0.233\textwidth]{figures/original-chinese.jpg}}
\subfigure[]{\includegraphics[width=0.233\textwidth]{figures/preprocess.png}}
\subfigure[]{\includegraphics[width=0.233\textwidth]{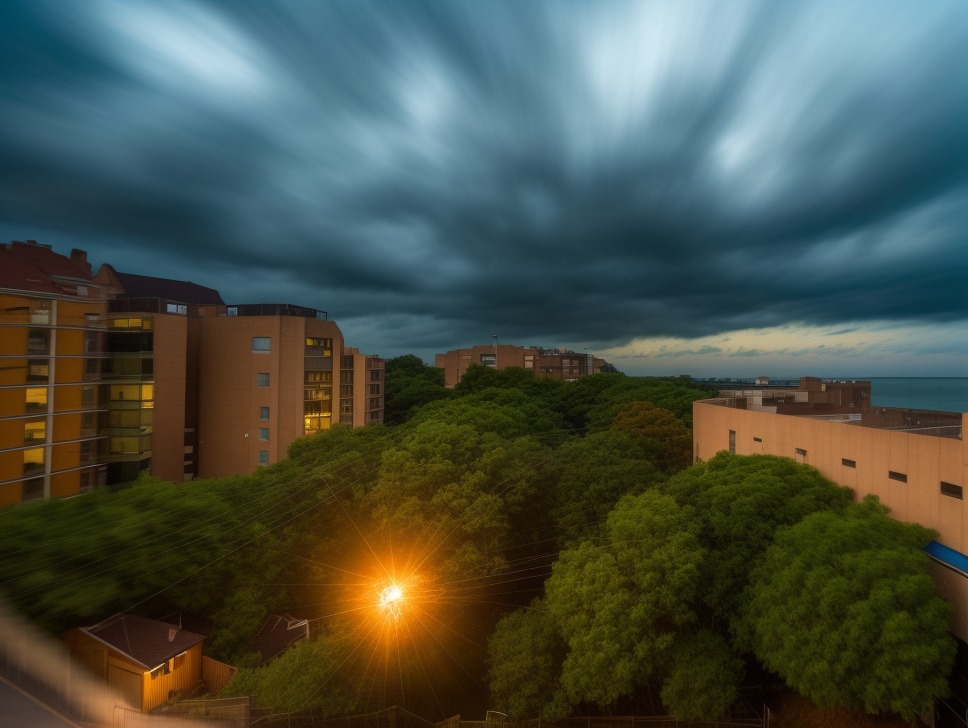}}
\subfigure[]{\includegraphics[width=0.233\textwidth]{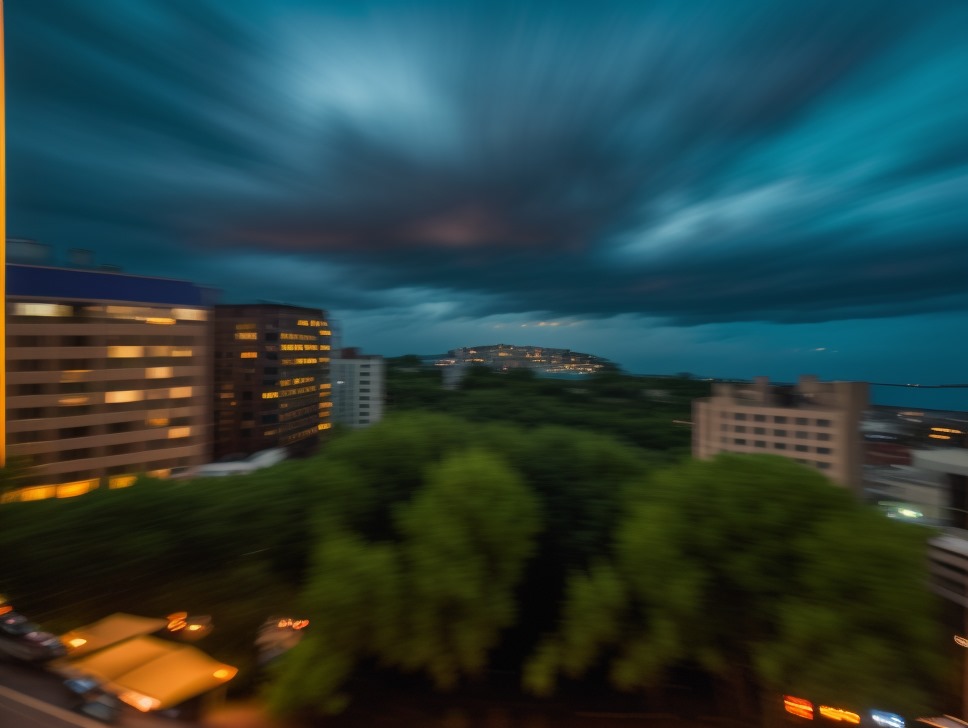}}
\caption{(a) shows the original monochrome image (324x244) provided as an input to the OneFormer (Coco) Segmentation Model (ControlNet), which extracts the location and shape of the objects, shown in (b) where we combine acceleration exhibited as (c) light motion blur (acceleration = 1.2 $m/s^2$) \& (d) obvious motion blur (acceleration = 3.8 $m/s^2$) along with environmental sensor stream (Ambient Light: 3196 Lux, Temperature: 28.04 $^{\circ}C$; Humidity: 71.54\%, Pressure: 1003.64 hPa, and Wind Velocity: 0.0 $m/s$) as an automated prompt generated by GPT-4 to the stable diffusion model to generate (c) a realistic image (968x728), emphasizing a dynamic and active environment as an output.}
\label{fig:E5}
\end{figure*}

\begin{figure*}[ht!]
\centering     
\subfigure[]{\includegraphics[width=0.243\textwidth]{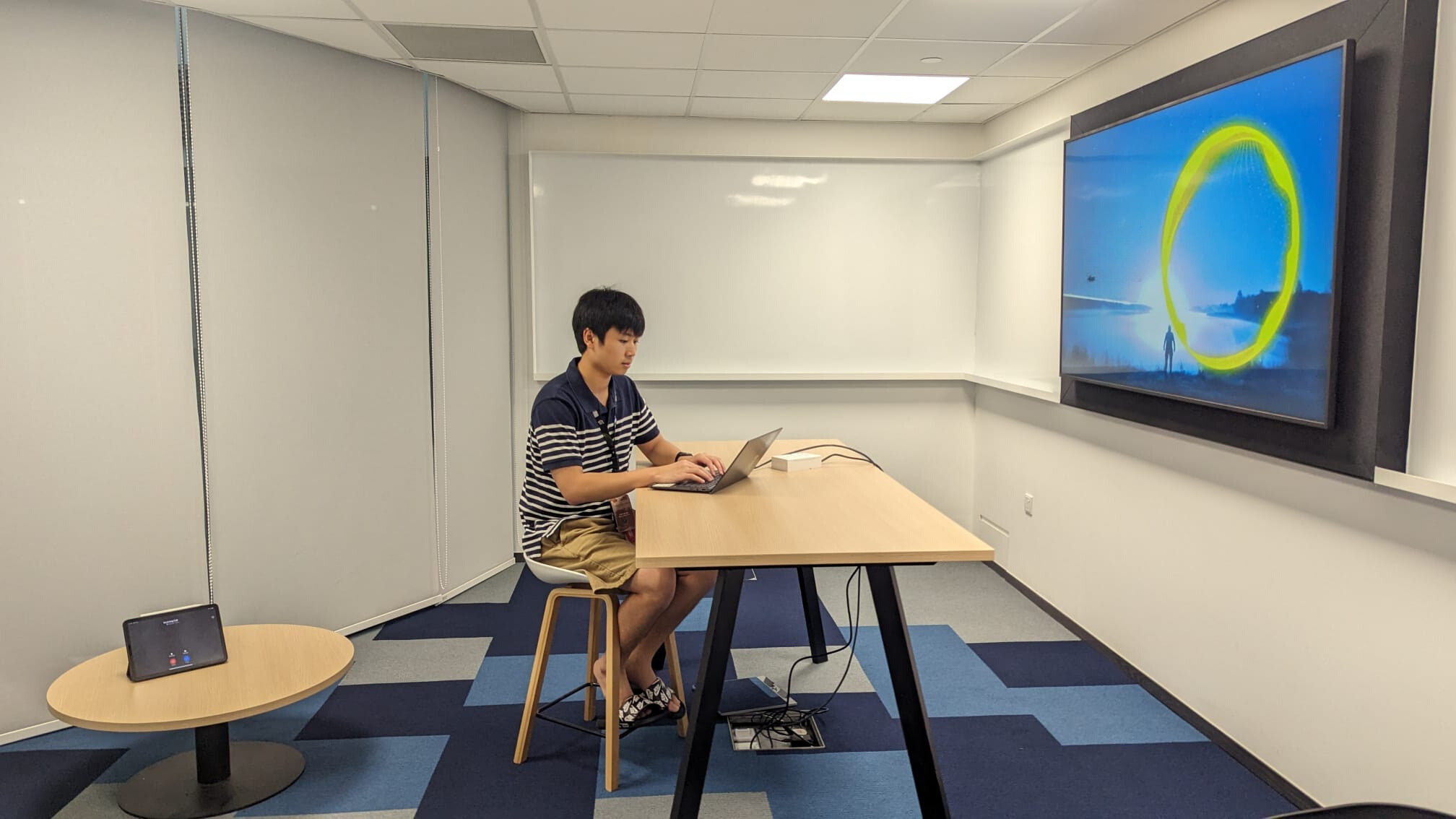}}
\subfigure[]{\includegraphics[width=0.243\textwidth]{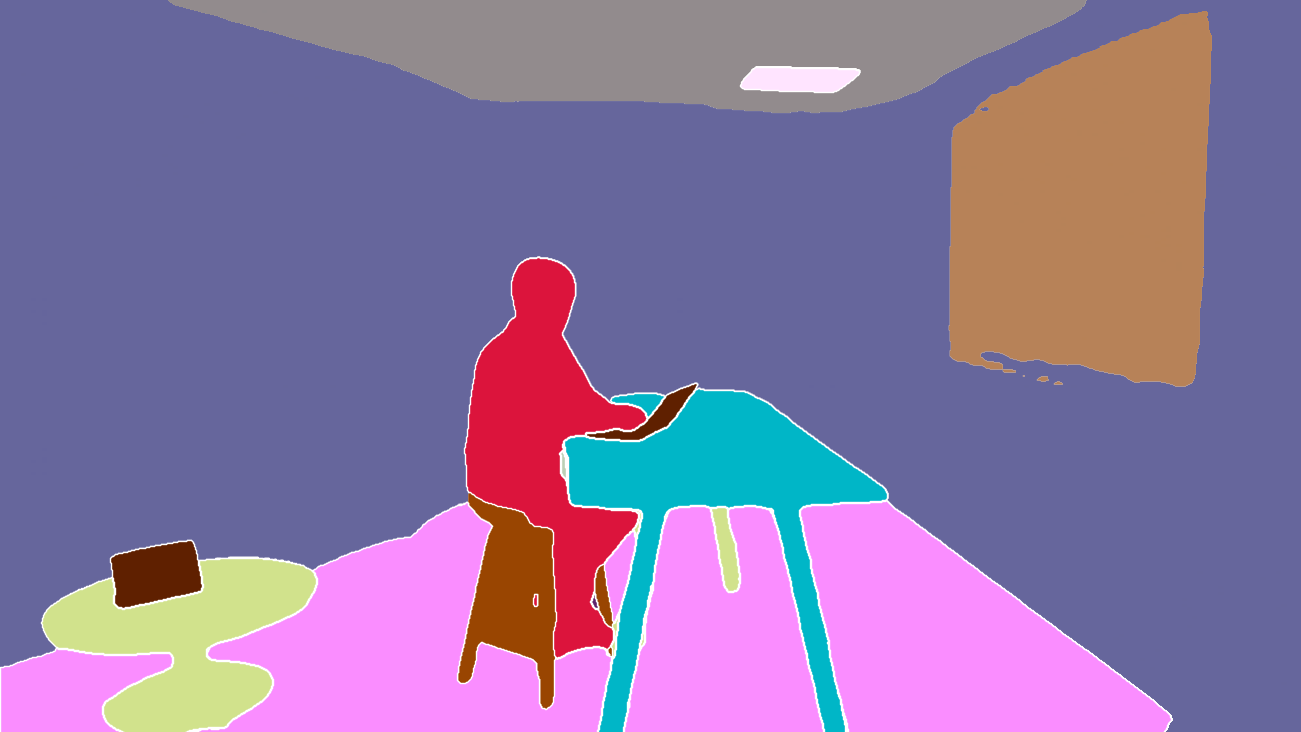}}
\subfigure[]{\includegraphics[width=0.243\textwidth]{figures/coloredcirclesonly.png}}
\subfigure[]{\includegraphics[width=0.243\textwidth]{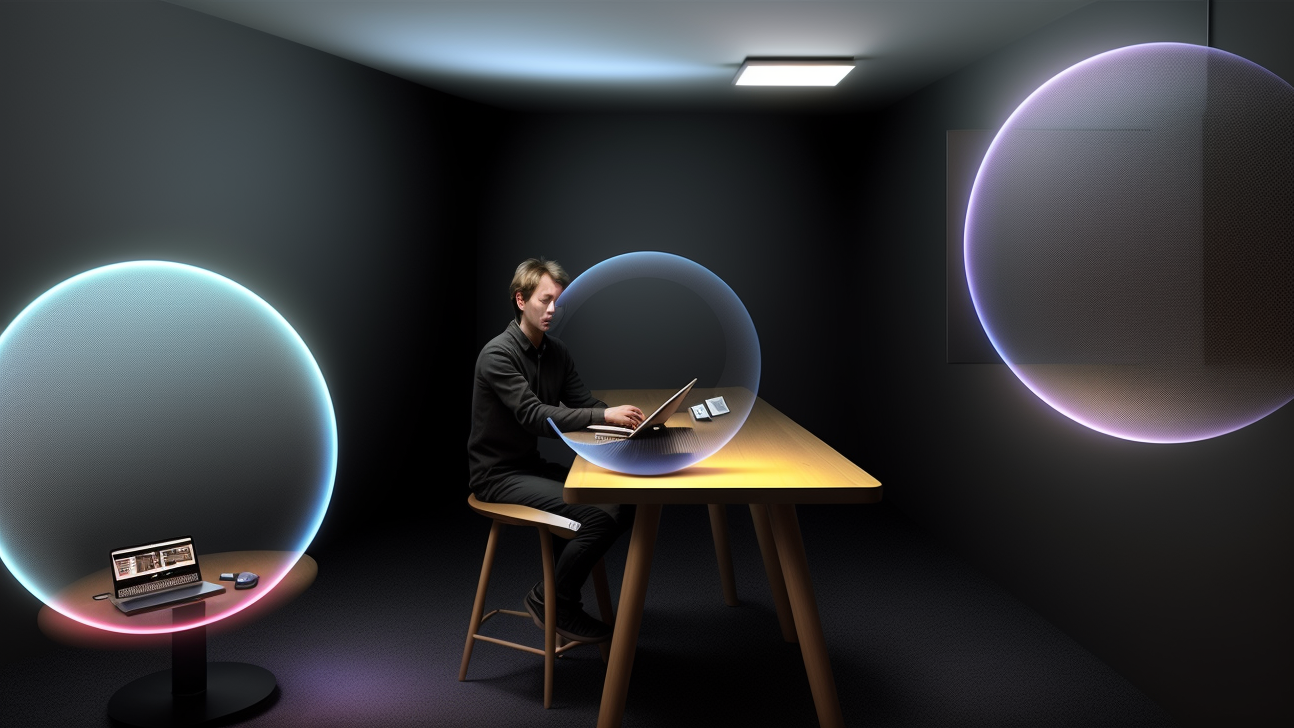}}
\\
\caption{(a) shows the original RGB image (4032x2268) provided as an input to the OneFormer (Coco) Segmentation Model (ControlNet), which extracts the location and shape of the objects shown in (b). In addition, we extract acoustic emissions using GPT-4 as a Segmentation Model, exhibited in (c) as large yellow circles, providing microphone sensor data corresponding to each location in the RGB image. Finally, we combine the ControlNet output and acoustic emission’s segmentation output along with the environmental sensor stream to the stable diffusion model to generate (d) a realistic image (1296x728) as an output.}
\label{fig:E6}
\end{figure*}

\begin{figure*}[h]
\centering     
\subfigure[]{\includegraphics[width=0.2\textwidth]{figures/original-chinese.jpg}}
\subfigure[]{\includegraphics[width=0.2\textwidth]{figures/preprocess.png}}
\\
\subfigure[]{\includegraphics[width=0.19\textwidth]{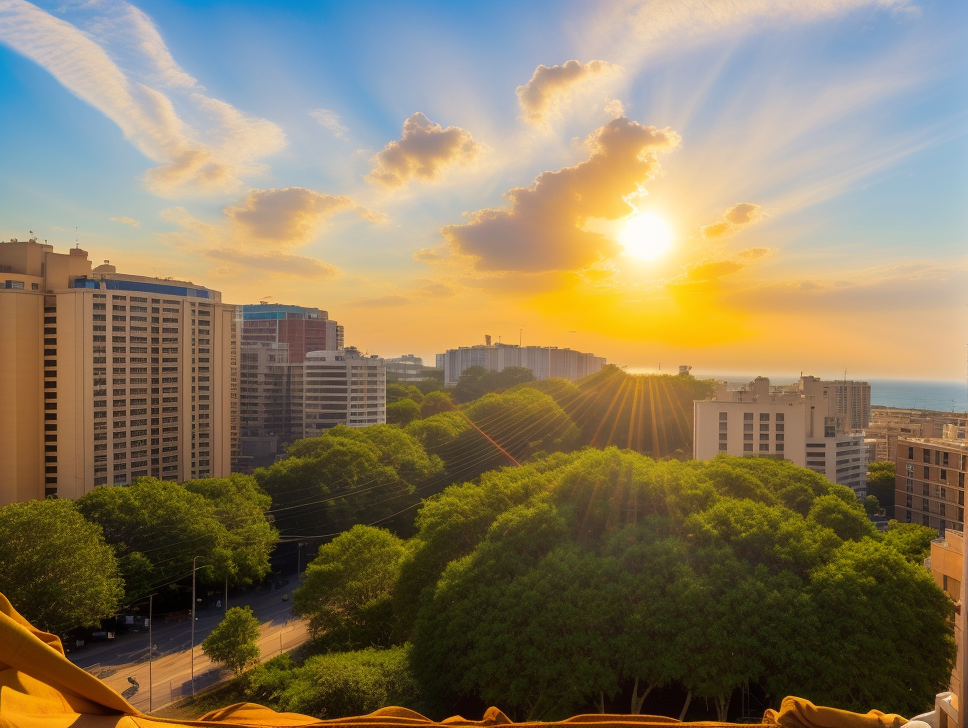}}
\subfigure[]{\includegraphics[width=0.19\textwidth]{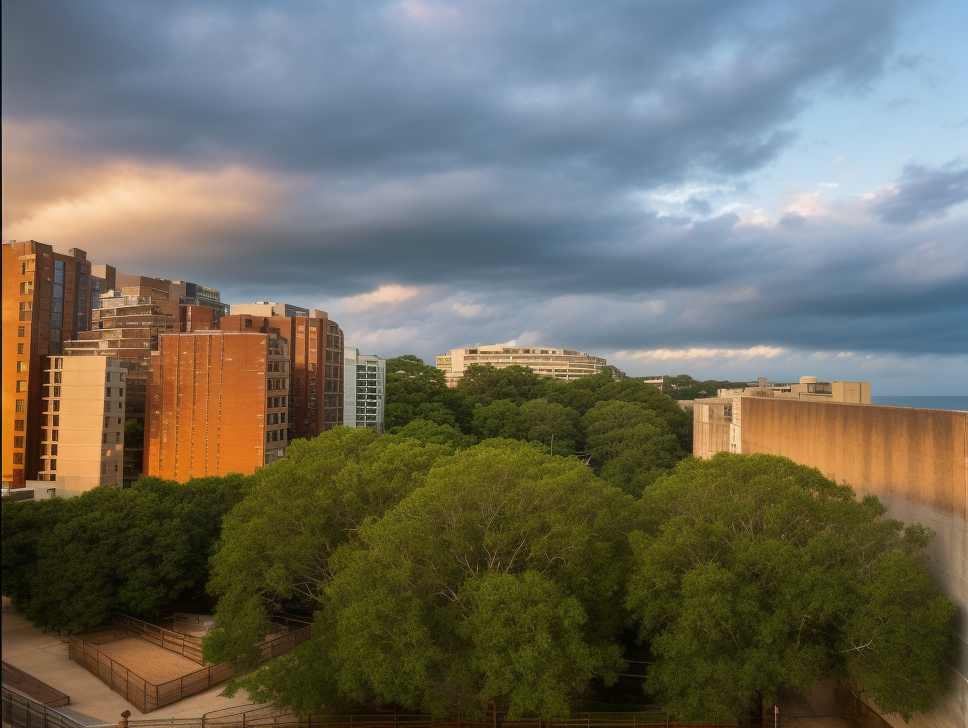}}
\subfigure[]{\includegraphics[width=0.19\textwidth]{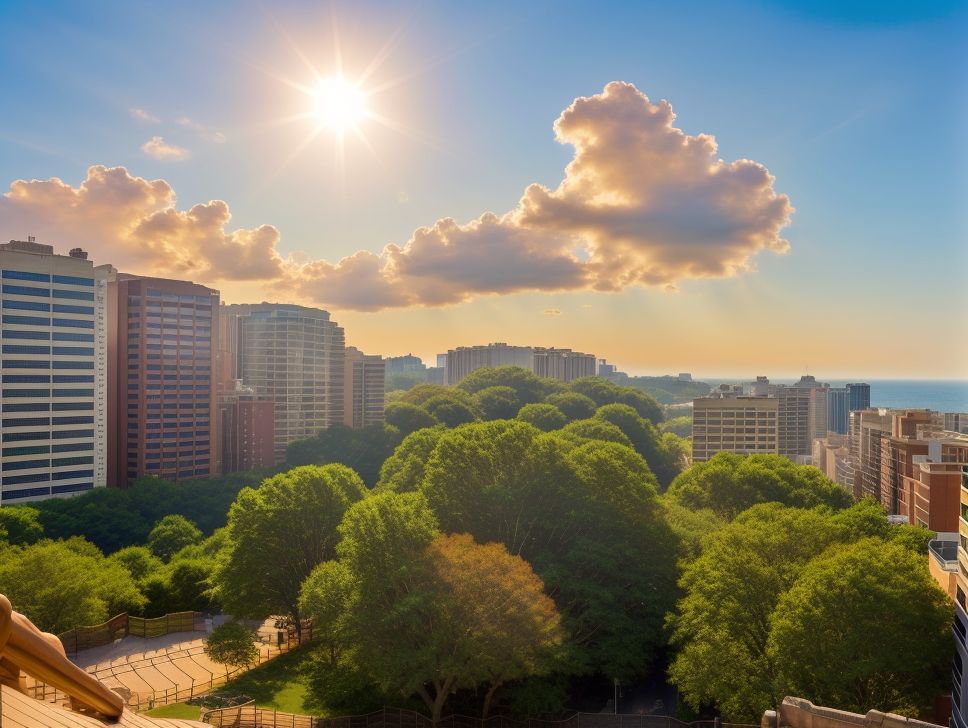}}
\subfigure[]{\includegraphics[width=0.19\textwidth]{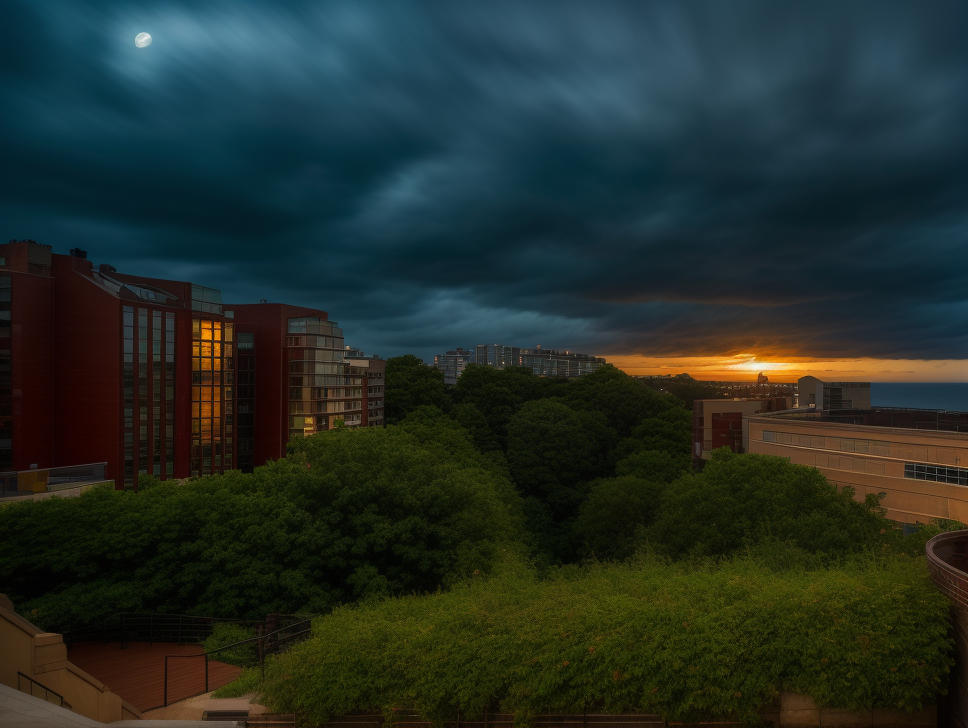}}
\subfigure[]{\includegraphics[width=0.19\textwidth]{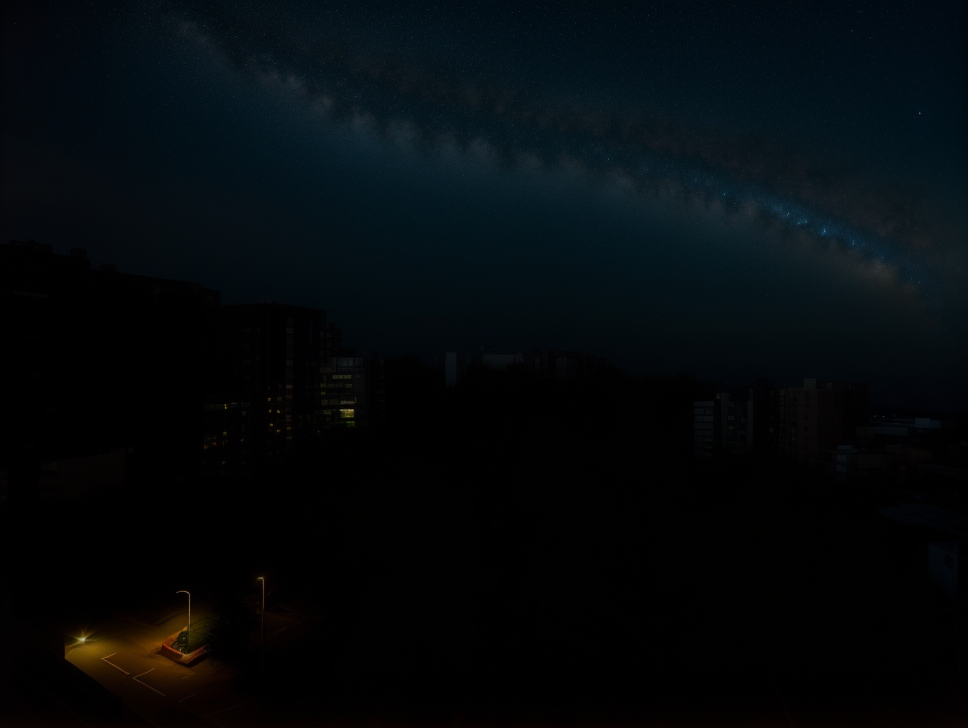}}
\caption{(a) shows the original monochrome image (324x244) provided as an input to the OneFormer (Coco) Segmentation Model (ControlNet), which extracts the location and shape of the objects, shown in (b), which we combine with the time-varying environmental sensor stream, fed as an automated prompt (shown in \ref{tab: video}) generated by GPT-4 to the stable diffusion model to generate (c-g) realistic images (968x728) as an output corresponding to different sensor values, e.g., (c) Lux = 65535, (d) Lux = 3089, (e) Lux = 16556, (f) Lux = 536, and (g) Lux = 0, while keeping segmentation output (background image) constant.}
\label{fig:E7}
\end{figure*}
\fakepar{Visualising motion} One of the capabilities of \system\space is its ability to visualize fields and phenomena that may be invisible to conventional cameras. In this specific experiment, we explore the visualization of motion as detected by the \hardware\space hardware. More precisely, we focus on visualizing acceleration that sensors can capture onboard the \hardware\space hardware. Our chosen method represents these time-varying phenomena as image artefacts, making them perceptible to the human eye in the generated image. More specifically, we demonstrate acceleration as motion blur in our visualizations. We generate suitable prompts using a LLM to achieve this, combining real-world monochrome images with custom acceleration data. Contrary to existing cameras, where blur maybe considered harmful, \system\space purposefully introduces blur to visualise acceleration.

We demonstrate the results of the experiment in the Figure~\ref{fig:E5}. The Figure~\ref{fig:E5}(a) shows the original captured monochrome image, the Figure~~\ref{fig:E5}(b) shows the output after the COCO model pre-processing, and the Figure~\ref{fig:E5}(c) \& (d) show the result of the image model corresponding to two different custom acceleration data, i.e., light motion~(acceleration, 1.2 $m/s^2$) \& obvious motion~(acceleration, 3.8 $m/s^2$). In the generated image, we show a significant contrast in the blur to denote the acceleration magnitude captured by \hardware\space.\\
\fakepar{Visualising acoustic emissions} We explore the ability of our system to visualise acoustic emissions in the room. This experiment combines several capabilities in the design of \system\space system. We deploy several \hardware\space in the environment to capture the acoustic emissions in an environment. We take an image of the environment and the captured acoustic emissions and provide these as input to a multimodal LLM. It helps us create a segmentation map of the environment, as shown in Figure~\ref{fig:E6}(b), thus localising various sources of acoustic emissions. Next, we use it with the image model to visualise the emissions as orange circles with varying sizes based on the intensity, as shown in Figure~\ref{fig:E6}(c). Finally, Figure~\ref{fig:E6}(d) demonstrates the  image generated by overlaying the image from the environment, with the image visualising acoustic emissions.

\begin{table*}[]
\begin{tabular}{|c|l|l|} \hline
Image & \multicolumn{1}{c|}{Sensor Data}& \multicolumn{1}{c|}{Prompt} \\ \hline
a     & \begin{tabular}[c]{@{}l@{}}Ambient Light: 65535 Lux\\ Temperature: 34.51 Celsius\\ Humidity: 63.14 \%\\ Pressure: 1006.42 hPa\\ Wind Velocity: 0.4 m/s\end{tabular} & \begin{tabular}[c]{@{}l@{}}Ultra-high-resolution photograph capturing an intensely sunny and hot day, \\ with clear skies, moderate humidity, and a calm atmosphere. The scene is \\ bathed in bright sunlight, creating deep shadows and vivid contrasts. Ideal for\\  8k UHD, high dynamic range imagery, shot with a Canon EOS-1D X Mark III.\end{tabular}              \\ \hline
b     & \begin{tabular}[c]{@{}l@{}}Ambient Light: 3089 Lux\\ Temperature: 27.42 Celsius\\ Humidity: 82.03 \%\\ Pressure: 1004.32 hPa\\ Wind Velocity: 2.6 m/s\end{tabular}  & \begin{tabular}[c]{@{}l@{}}High-resolution photo capturing a warm and overcast day with high humidity, \\ creating a slightly muggy feel. The sky is cloud-covered, diffusing the light and \\ casting a soft glow. A noticeable breeze ripples through the scene. Ideal for 8k \\ UHD, captured with a high dynamic range, using a Leica SL2.\end{tabular}                \\ \hline
c     & \begin{tabular}[c]{@{}l@{}}Ambient Light: 16556 Lux\\ Temperature: 30.69 Celsius\\ Humidity: 67.07 \%\\ Pressure: 1003.70 hPa\\ Wind Velocity: 0.0 m/s\end{tabular} & \begin{tabular}[c]{@{}l@{}}Ultra-high-resolution image of a bright, sunny day with clear skies, warm \\ temperatures, and moderate humidity. The sun casts sharp, defined shadows, and \\ the calm, windless atmosphere creates a serene outdoor scene. Perfect for 8k UHD, \\ high dynamic range photography, shot with a Sony Alpha A9.\end{tabular}                     \\ \hline
d     & \begin{tabular}[c]{@{}l@{}}Ambient Light: 536 Lux\\ Temperature: 27.79 Celsius\\ Humidity: 73.83 \%\\ Pressure: 1003.53 hPa\\ Wind Velocity: 1.5 m/s\end{tabular}   & \begin{tabular}[c]{@{}l@{}}High-quality image depicting a warm evening with a heavily overcast sky, creating \\ a dim and moody atmosphere. The air is humid and slightly breezy, adding to the \\ sense of an impending nightfall. Ideal for 8k UHD, high dynamic range, captured \\ with a Nikon D850.\end{tabular}                                                      \\ \hline
e     & \begin{tabular}[c]{@{}l@{}}Ambient Light: 0 Lux\\ Temperature: 26.85 Celsius\\ Humidity: 79.26 \%\\ Pressure: 1004.63 hPa\\ Wind Velocity: 0.0 m/s\end{tabular}     & \begin{tabular}[c]{@{}l@{}}High-resolution night photograph capturing a warm and very humid environment \\ in complete darkness, conveying a still and heavy night atmosphere. The air is thick \\ with moisture, creating a sense of quiet and solitude. Ideal for 8k UHD, high dynamic \\ range, shot with a Canon EOS R6, excellent for night photography.\end{tabular} \\ \hline
\end{tabular}
\caption{shows the image index corresponding to the stream of realistic images (Figure~\ref{fig:E7}) along with the actual time-varying environmental sensor data (shown in the second column) and the corresponding positive prompt generated using GPT-4 in the third column, which is fed to the diffusion model to produce output.}
\label{tab: video}
\end{table*}
\fakepar{Generating image sequences} Finally, we investigate the ability of \system\space to generate a sequence of images. 
Till now, we only show how \name\space system can generate creative and novel image representations. However, it can also be a stack of images, essentially put together as a video. To enable this, we trigger a unique (different) prompt for each sensor data; hence, the corresponding output also varies. We present some of the keyframes from this analysis in Figure~\ref{fig:E7}. Here is the link~\cite{video_url} to the video output (combined stream of images).
We observe that the sunlight (ambient light intensity) varies significantly from Figure~\ref{fig:E7}(c) to \ref{fig:E7}(g). Also, the cloud density varies along with this. At the same time, the composition of the actual image (similar to the original monochrome image) is intact. We also present the list of positive prompts generated using GPT-4, which we use to generate these realistic images along with original environmental sensor data in Table~\ref{tab: video}.
\section{Discussion}

\fakepar{Power consumption} \hardwareF\space platform hosts Apollo3 Blue SOC as the main MCU and several sensors like a monochrome image sensor, environmental sensor, light sensor, acoustic sensor, and IMU sensor. All of them support active mode as well as low-power mode  consuming  power consumption in hundreds of microwatts to few-milliwatts. This is significantly lower when compared to high resolution ECS.

\fakepar{Bandwidth } \system\space system consumes very little bandwidth regarding image and sensor data acquisition and fairly optimizes our net bandwidth utilization compared to an ECS system capable of delivering a similar-sized RGB image. We acquire one image with a 324x244 resolution, i.e., 78,568 pixels with a bit depth of 8 bits, accounting for 628,544 bits per image. Likewise, we acquire light sensor data, i.e., 16 bits, and temperature, humidity, and pressure data (20 bits each), i.e., 60 bits. We also acquire 14 bits of data from the audio sensor using on-chip ADC and 48 bits from the accelerometer (16 bits corresponding to each axis). ollectively, we acquire approximately 628,682 bits (or 78.585kBs) of data and generate an HD RGB image with a resolution of 968x728, i.e., 704704 pixels. The generated RGB image has a bit depth of 24 bits, accounting for 16,912,896 bits (or 2.114MBs). Comparatively, our system utilizes around 27 times less bandwidth than a traditional camera system with the given resolution. This leads to lower data being sent over the air, and hence, lowered power consumption.




\fakepar{Prompting and hallucinations}  Image and language models often encounter issues with hallucinations, and \system is no exception. We have observed that \system may produce varying representations for similar prompts. Additionally, the quality of the generated images is influenced by the sophistication and quality of the prompt. Nonetheless, many of these challenges can be mitigated through fine-tuning and other ongoing advancements.

\fakepar{Prompt}
Diffusion models typically require a text input as a prompt to generate an image. In addition, they need control information in the form of a negative prompt to produce realistic images and avoid bad images as output.
Here, we provide a sample scheme to generate these prompts, manually leveraging environment data captured using sensors on board.
\vspace{5pt}
\begin{tcolorbox}[colback=white,colframe=black,arc=0pt,outer arc=0pt,boxrule=0.5pt,title=Prompt,fontupper=\ttfamily\normalsize,breakable]
Please provide a unified description of the attached monochrome image, integrating the following sensor data: (replace a, b, c,d with sensor data collected)\\
"""\\
Ambient light intensity: a Lux,\\
Temperature: b Celsius,\\
Humidity: c Percentage,\\
Pressure: d Pa.\\
"""\\
\end{tcolorbox}
Note that we can use additional sensor information with this Prompt to augment the image, like depth information. 
Also, we can ask multimodal models to write Prompts to diffusion models to generate creative images based on the user's preference.


\section{Conclusion}
We have introduced \system, a rethink of  the architecture of modern cameras, made possible by the convergence of advancements in various fields, including low-power sensing combined with language and vision models. \system, in contrast to traditional cameras, captures a rich representation of the world, which is then visualized using image models. For instance, it can collect low-bitrate environmental data using simple sensors and a monochrome camera yet generate high-resolution, bandwidth-intensive images at the edge device. This capability paves the way for developing low-power embedded hardware capable of streaming high-resolution images while operating on small batteries or harvested energy. Furthermore, \system\space can visualize fields and phenomena typically invisible to conventional cameras, as demonstrated through an extended reality headset. We believe that \system\space opens new avenues for research and photography.

\section{Acknowledgements}

We thank the anonymous reviewers for their insightful comments. This work is funded through a grant from the NUS-NCS Center~(A-0008542-02-00) that is hosted at the National University of Singapore. This work is also supported through a startup grant~(A-8000277-00-00)  from the National University of Singapore.

\bibliographystyle{ACM-Reference-Format}
\bibliography{sample}

\end{document}